\documentclass[acmtog]{acmart}
\acmSubmissionID{1203}

\usepackage{diagbox}
\usepackage{verbatim}

\usepackage{amssymb}
\usepackage{latexsym}
\usepackage{lineno}
\usepackage{xcolor}
\usepackage{tabularx}
\usepackage{xspace}
\usepackage{color}
\graphicspath{{figures/}}
\usepackage{amsmath,bm}
\usepackage{psfrag}
\usepackage{mathtools}
\usepackage{hyperref}

\usepackage{multirow, tabularx}
\newcolumntype{Y}{>{\centering\arraybackslash}X}
\usepackage{capt-of}%
\usepackage{array, boldline, makecell, booktabs}

\usepackage{array}

\usepackage{pifont}

\usepackage{wrapfig}
\usepackage{psfrag}
\usepackage{makecell}
\usepackage{physics}
\usepackage{algorithm}
\usepackage{threeparttable,booktabs}
\usepackage{graphicx}
\usepackage{subcaption}
\usepackage{flushend}
\usepackage{algpseudocode}
\usepackage{fancyhdr}
\usepackage{placeins}
\AtBeginDocument{%
  }

\newcommand{\figvspace}{\vspace{-0.2cm}}

\AtBeginDocument{%
  }

\setcopyright{rightsretained}
\acmJournal{TOG}
\acmYear{2024} \acmVolume{43} \acmNumber{6} \acmArticle{} \acmMonth{12}\acmDOI{10.1145/3687996}

\begin{document}

\author{Sinan Wang}
\email{swang3081@gatech.edu}
\affiliation{
\institution{Georgia Institute of Technology}
\country{USA}
}
\affiliation{
  \institution{The University of Hong Kong}
  \country{Hong Kong}
}

\author{Yitong Deng}
\email{yitongd@stanford.edu}
\affiliation{
\institution{Stanford University}
\country{USA}
}

\author{Molin Deng}
\email{mdeng47@gatech.edu}
\affiliation{
\institution{Georgia Institute of Technology}
\country{USA}
}

\author{Hong-Xing Yu}
\email{koven@cs.stanford.edu}
\affiliation{
\institution{Stanford University}
\country{USA}
}

\author{Junwei Zhou}
\email{zjw330501@gmail.com}
\affiliation{
\institution{Purdue University}
\country{USA}
}
\affiliation{
\institution{University of Michigan}
\country{USA}
}

\author{Duowen Chen}
\email{dchen322@gatech.edu}
\affiliation{
\institution{Georgia Institute of Technology}
\country{USA}
}

\author{Taku Komura}
\email{taku@cs.hku.hk}
\affiliation{
\institution{The University of Hong Kong}
\country{Hong Kong}
}

\author{Jiajun Wu}
\email{jiajunwu.cs@gmail.com}
\affiliation{
\institution{Stanford University}
\country{USA}
}

\author{Bo Zhu}
\email{bo.zhu@gatech.edu}
\affiliation{
\institution{Georgia Institute of Technology}
\country{USA}
}

\title{An Eulerian Vortex Method on Flow Maps}

\begin{abstract}
We present an Eulerian vortex method based on the theory of flow maps to simulate the complex vortical motions of incompressible fluids. Central to our method is the novel incorporation of the flow-map transport equations for \emph{line elements}, which, in combination with a bi-directional marching scheme for flow maps, enables the high-fidelity Eulerian advection of vorticity variables. The fundamental motivation is that, compared to impulse $\bm m$, which has been recently bridged with flow maps to encouraging results, vorticity $\bm \omega$ promises to be preferable for its numerical stability and physical interpretability. To realize the full potential of this novel formulation, we develop a new Poisson solving scheme for vorticity-to-velocity reconstruction that is both efficient and able to accurately handle the coupling near solid boundaries. We demonstrate the efficacy of our approach with a range of vortex simulation examples, including leapfrog vortices, vortex collisions, cavity flow, and the formation of complex vortical structures due to solid-fluid interactions.
\end{abstract}



\begin{teaserfigure}
 \centering
 \includegraphics[width=0.995\textwidth]{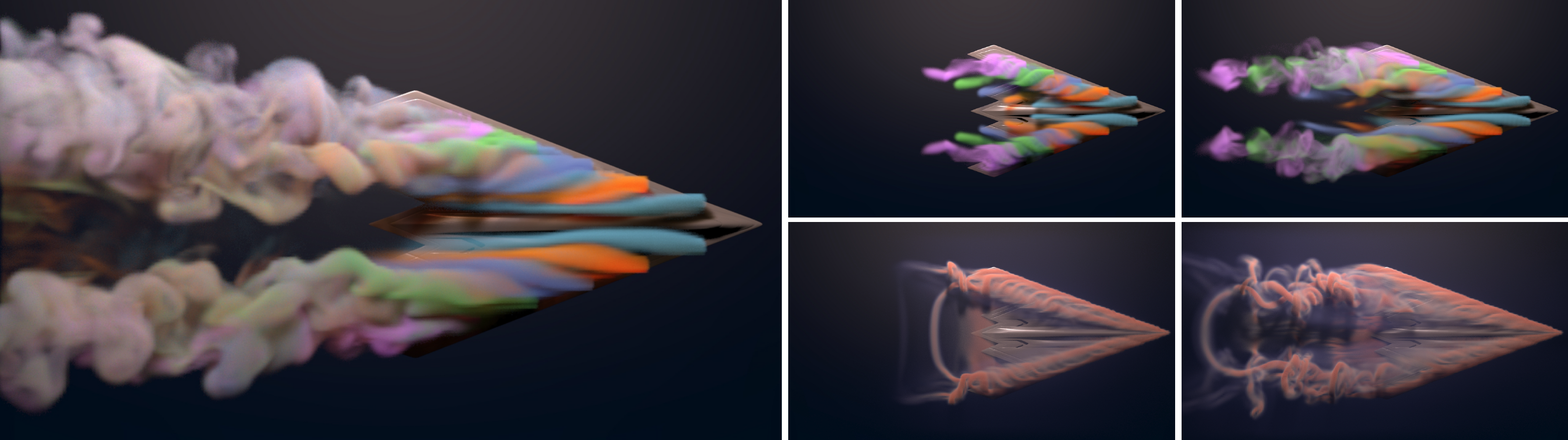}
 \figvspace
\caption{Turbulent flow generated from a delta wing with zero viscosity and an angle-of-attack of \(20^\circ\). Two pictures on the right illustrate smoke (top) and vorticity (bottom) snapshots at different time steps. A more stable spiral vortex structure along the two leading edges can be generated when viscosity is considered, as shown in Figure~\ref{fig:ray_and_delta_viscosity_comp} (top).}
\label{fig:delta_wing}
\end{teaserfigure}

\maketitle

\vspace{0.1in}
\section{Introduction}
Vortex methods have established themselves as a foundational computational method for incompressible flows with turbulence and complex vortical structures. In these methods, the Navier-Stokes equations written in the velocity form $\bm u$ are reformulated into their vorticity form, establishing vorticity
\begin{equation}
    \bm \omega = \nabla \times \bm u,
    \label{eq: vort_compatibility}
\end{equation}
as the primary physical quantity for fluid representation. Such a vorticity-based formulation is favored by researchers in both theoretical and experimental fluid dynamics for its unambiguous physical meaning, distinctive topological structures, and direct correlation with visually appealing vortical details. Over the past decade, extensive research effort across computational physics and computer graphics has been devoted to advancing vorticity-based incompressible flow solvers (see the work by \citet{mimeau2021review} for a comprehensive survey), exploring a rich set of data structures including particles \cite{selle2005vortex,cottet2000vortex, park2005vortex,zhang2014pppm,angelidis2017multi}, filaments \cite{angelidis2005simulation, weissmann2010filament, padilla2019bubble, ishida2022hidden}, sheets \cite{brochu2012linear, pfaff2012lagrangian, da2015double}, as well as their hybridization \cite{chern2016schrodinger, chern2017inside, yang2021clebsch, xiong2022clebsch}. 

Despite these innovative advancements, it stands as a peculiar (yet reasonable) fact that pure Eulerian vortex methods are very rare. Unlike their velocity-based counterparts, which benefit from the parallelizable and large-scale solver-friendly nature of Cartesian grids (e.g., the stable fluid solver \cite{stam1999stable}), vortex methods typically require specific auxiliary data structures, usually particles or other forms of Lagrangian discretizations, to facilitate their temporal evolution. Particularly, owing to the challenges in robustly handling the vorticity stretching term and preserving both the structure and amount of vorticity during its advection, it is difficult to devise a numerical scheme that evolves vorticity on Cartesian grids in a purely Eulerian manner by directly adopting off-the-shelf advection schemes like Semi-Lagrangian \cite{sawyer1963semi, jameson1981numerical} or BFECC \cite{kim2006advections, selle2008unconditionally}, as in these cases the vorticity representation offers no significant advantages over velocity in capturing the fluid's vortical motion, which renders these efforts less appealing than the conventional approaches, particularly in the computer graphics community.

To this end, we propose a novel Eulerian vortex method tailored for graphical simulation applications. Our framework is based on the mathematical insight that the vorticity-based fluid equations can be elegantly and accurately solved using the transport equations for \textit{line elements} on long-range flow maps, echoing recent advances in the impulse-based fluid methods in computer graphics \cite{nabizadeh2022covector,deng2023fluid}, which have achieved state-of-the-art simulation accuracy through combining long-range flow maps with the \textit{surface element} impulse. However, despite the encouraging results, a surface element like impulse is inherently limited by its tendency to rapidly and constantly increase in magnitude, which destabilizes the simulation. In comparison, the line element vorticity promises to offer a more advantageous pairing for being numerically stable and physically interpretable, offering direct connections to the fluid's vortical structures, while enjoying the same low-dissipation advantages as impulse-based methods. 

Such an observation motivates the design of our Eulerian vortex method that evolves the vorticity $\bm \omega$ based on long-range flow maps computed on Cartesian grids. First, we employ a bi-directional marching mechanism to evolve accurate flow maps and their Jacobians. Then, we employ an error-compensated advection scheme to compute our novel \textit{line element vorticity advection} based on the computed flow maps. Finally, we devise a novel Poisson solving scheme to reconstruct velocity $\bm u$ from our evolved vorticity $\bm \omega$. For such reconstruction, standard methods \cite{Ando:2015:streamfunc, elcott2007stable, 10.1145/3592402, zhang2015restoring} utilize the vorticity-streamfunction formulation, which requires additional velocity potential solving in order to handle the solid boundary conditions. In comparison, we propose an alternative solution based on the velocity-vorticity formulation, where, instead of the streamfunction, the velocity serves as the unknown of a Poisson system, allowing for straightforward incorporation and direct enforcement of solid boundary conditions at the solver level. As discussed in \cite{cottet2000vortex}, it is possible that by means of Poincaré identity, we can enforce the solid boundary conditions by directly considering the relationship between the vorticity and the velocity field, thus bypassing the need for the streamfunction and potential. Drawing on this conceptual insight, we implement a novel velocity-vorticity solution using an efficient, matrix-free GPU-based Poisson solver to solve the system.
We verify the correctness, versatility, and efficacy of our approach with a diverse set of challenging vortex simulation scenarios, including vortex shedding and development from moving solids, leapfrogging vortices, as well as vortex collision and reconnections.

\section{Related Work}

\paragraph{Grid-based fluid simulation}
Since the seminal work of \citet{stam1999stable}, grid-based solvers have been used to simulate a wide variety of physical phenomena \cite{fedkiw2001visual, foster2001practical, nguyen2002physically}. In addition to regular Cartesian grids, sparse and adaptive grid structures like Octree \cite{rasmussen2004directable, losasso2004simulating, losasso2006spatially, aanjaneya2017power, ando2020practical}, SPGrid \cite{setaluri2014spgrid}, Multi-grid \cite{mcadams2010parallel}, RLE grid \citep{irving2006efficient, houston2006hierarchical, chentanez2011real} and far-field grid \citep{zhu2013new} have also been introduced to increase the effective resolution and the size of the simulation domain, as well as to reduce numerical dissipation.

\paragraph{Flow map method}
Flow map advection, also referred to as the method of characteristic mapping (MCM), was first adopted for fluid simulation by \citet{wiggert1976numerical}, and later introduced to the computer graphics community by \citet{tessendorf2011characteristic}. Virtual particles are typically used to compute flow maps \cite{hachisuka2005combined, sato2017long, sato2018spatially, tessendorf2015advection}, while \citet{qu2019efficient} proposed a Semi-Lagrangian-like scheme to advect flow maps in the Eulerian manner to mitigate the time cost. Such methods significantly reduce numerical dissipation. Later,
\citet{nabizadeh2022covector} and \citet{deng2023fluid} combined the flow map with gauge fluid methods using impulse variables and verified the crucial role that the flow map plays in the accurate advection of these variables. Recently, \citet{li2024lagrangian} advanced this approach by developing a Lagrangian method that successfully handles the free surface case. We extend this to the vorticity-velocity Navier-Stokes equations, which is another form of gauge applied to the usual velocity-pressure form \cite{mercier2020characteristic, yin2021characteristic, yin2023characteristic}.

\paragraph{Vortex method}
Vortex methods rewrite the incompressible Navier-Stokes equation by treating vorticity as a gauge variable for velocity. By explicitly advecting the vorticity, this approach naturally preserves fluid circulation. In order to reduce the numerical dissipation during the advection, diverse vorticity representations have been proposed, ranging from particles \cite{selle2005vortex,cottet2000vortex,park2005vortex, zhang2014pppm,angelidis2017multi}, to filaments \cite{angelidis2005simulation, weissmann2010filament, ishida2022hidden, padilla2019bubble}, segments \cite{xiong2021incompressible}, sheets \cite{brochu2012linear, pfaff2012lagrangian, da2015double}, and Clebsch level sets \cite{chern2016schrodinger, chern2017inside, yang2021clebsch, xiong2022clebsch}. For these Lagaragian vorticity representations, the velocity reconstruction from vorticity is often done with the Biot-Savart law, accelerated with the Fast Multipole Method (FMM) \cite{greengard1987fast}. Despite the enhanced ability to preserve the vortex structure, these vorticity representations are not as convenient to implement as the Eulerian vortex method when handling solid boundaries. Vortex methods with the vorticity-streamfunction formulation have been adopted by several researchers \cite{zhang2015restoring, elcott2007stable, Ando:2015:streamfunc, 10.1145/3592402}. In spite of the ability to handle solid boundaries, these methods require solving a potential component of the velocity. Existing Eulerian vortex methods that use the velocity-vorticity formulation avoid the streamfunction but suffer from poor vorticity preservation and inaccurate solid boundary conditions. For example, \citet{huang1997finite} only addressed 2D cases and \citet{liu2001numerical} set the vorticity boundary condition simply as \(\bm{\omega}_b = 0\).
 Our method bypasses the use of streamfunction and potential without sacrificing the correctness in solid boundary handling, and pushes forward the state-of-the-art in vortex method simulations through the novel combination with flow map advection.

\section{Physical Model}
\label{section: physical_model}

\paragraph{Naming Convention}
In this paper, all subscripts are used to indicate axes (like \(x\), \(y\), \(z\)) while superscripts are used for cell or face indices (like \(i\), \(j\), \(k\)). All equations are presented in their matrix forms. We also present the definitions of main variables in Table~\ref{tab:notation_table}.

\subsection{Flow map}
The forward flow map, \(\bm{\phi}(\cdot, t)\), is defined as a function of space and time, mapping the initial position of a fluid particle at time \(0\) to its position at a subsequent time \(t\). Analogously, the backward flow map, \(\bm{\psi}(\cdot, t)\), is defined as the mapping from the position of a fluid particle at time \(t\) back to its original position at time \(0\). 

For the forward flow map \(\bm{\phi}\):
\begin{equation}
    \begin{dcases}
        \frac{\partial \bm{\phi}\left(\bm{X}, \tau\right)}{\partial \tau} = \bm{u}\left(\bm{\phi}\left(\bm{X}, \tau\right), \tau\right), \\
        \bm{\phi}\left(\bm{X}, 0\right) = \bm{X}, \\
        \bm{\phi}\left(\bm{X}, t\right) = \bm{x}.
    \end{dcases}
    \label{eq:phi_def}
\end{equation}

Its inverse, the backward flow map \(\bm{\psi}\) can then be defined as
\begin{equation}
    \begin{dcases}
        \bm{\psi}\left(\bm{x}, t\right) = \bm{x}, \\
        \bm{\psi}\left(\bm{x}, 0\right) = \bm{X}.
    \end{dcases}
    \label{eq:flow_map}
\end{equation}

In these definitions, \(\bm{X}\) denotes the position of a material point at time \(0\), while \(\bm{x}\) refers to the position of the same material point at time \(t\). 
The Jacobians of the functions \(\bm{\phi}\) and \(\bm{\psi}\) are denoted by \(\mathcal{F}\) and \(\mathcal{T}\) respectively: \(\mathcal{F} := \frac{\partial \bm{\phi}}{\partial \bm{X}}\), \(\mathcal{T} := \frac{\partial \bm{\psi}}{\partial \bm{x}}\).
In particular, \(\mathcal{F}\) is the forward Jacobian evaluated at the undeformed position \(\bm{X}\), while \(\mathcal{T}\) is the backward Jacobian evaluated at the deformed position \(\bm{x}\).

For a specific flow map defined on a moving particle, the evolution of the Jacobians \(\mathcal{F}\) and \(\mathcal{T}\) is governed by:
\begin{equation}
    \begin{aligned}
        \frac{D \mathcal{F}}{D t} &= \nabla\bm{u}\,\mathcal{F}, \label{eq:F_evolution} \\
        \frac{D \mathcal{T}}{D t} &= -\mathcal{T}\,\nabla\bm{u}.
    \end{aligned}
\end{equation}
Here, \(\frac{D(\cdot)}{Dt}\) represents the material derivative, which describes the rate of change of the Jacobians moving with the particle along its flow map's trajectory. 

\newcolumntype{z}{X}
\newcolumntype{s}{>{\hsize=.25\hsize}X}
\begin{table}
\caption{Summary of the main symbols and notations.}
\vspace{-0.1in}
\centering
\small
\begin{tabularx}{0.47\textwidth}{scz}
\hlineB{2.5}
Notation & Type & Definition\\
\hlineB{2.5}
\hspace{12pt}$\bm{X}$ & vector & material point position at initial state\\
\hline
\hspace{12pt}$\bm{x}$ & vector & material point position at time t\\
\hline
\hspace{12pt}$t$ & scalar & time\\
\hline
\hspace{12pt}$\tau$ & scalar & dummy variable\\
\hline
\hspace{12pt}$\bm{\phi}$ & vector & forward map\\
\hline
\hspace{12pt}$\bm{\psi}$ & vector & backward map\\
\hline
\hspace{12pt}$\mathcal{F}$ & matrix & forward map gradients\\
\hline
\hspace{12pt}$\mathcal{T}$ & matrix & backward map gradients\\
\hline
\hspace{12pt}$\mathcal{V}$ & vector & velocity buffer\\
\hline
\hspace{12pt}$\mathcal{S}$ & vector & time buffer\\
\hline
\hspace{12pt}$\bm{u}$ & vector & velocity\\
\hline
\hspace{12pt}$\bm{m}$ & vector & impulse\\
\hline
\hspace{12pt}$\bm{\omega}$ & vector & vorticity\\
\hline
\hspace{12pt}$\bm{s}$ & vector & surface element\\
\hline
\hspace{12pt}$\bm{l}$ & vector & line element\\
\hline
\hspace{12pt}$n$ & scalar & number of steps between reinitializations\\
\hlineB{2.5}
\end{tabularx}
\captionsetup{aboveskip=20pt}
\label{tab:notation_table}
\vspace{-0.3cm}
\end{table}

\paragraph{Line Elements and Surface Elements} 
Next, we define the flow map-based advection equations for line elements and surface elements , and we will show that vorticity is a line element and impulse is a surface element. Geometrically, line and surface elements amount to 2-forms and 1-forms in differential geometry (see \cite{nabizadeh2022covector} for detailed discussion).
In an incompressible flow field \(\bm u\), if the advection of a vector field \(\bm{l}\) satisfies 
\begin{equation}
    \frac{D \bm l}{Dt}=\left(\nabla \bm u\right) \bm l,
    \label{eq:line_adv}
\end{equation}
we define \(\bm l\) as a \emph{line element}.
The advection of a line element can be characterized by the bidirectional flow map 
\begin{equation}
    \bm{l}\left(\bm x, t\right) = \mathcal{F}\,\bm{l}\left(\bm{\psi}\left(\bm{x}\right), 0\right),
    \label{eq:line_ffm}
\end{equation}
\begin{equation}
    \bm{l}\left(\bm{X}, 0\right) = \mathcal{T}\,\bm{l}\left(\bm{\phi}\left(\bm{X}\right), t\right).
    \label{eq:line_bfm}
\end{equation}
Here, Eqn.~\eqref{eq:line_ffm} defines the forward evolution, and Eqn.~\eqref{eq:line_bfm} defines the time-reversed evolution. In Eqn~\eqref{eq:line_ffm}, we have \(\bm{\psi}\left(\bm{x}\right)\) and \(\bm{x}\) specifying the start and end point of the flow map path at time \(0\) and \(t\).
Conversely, in Eqn~\eqref{eq:line_bfm}, by examining the backward path from time \(t\) to time \(0\), we can construct a mapping with \(\bm{l}\left(\bm{X}, 0\right)\) denoting the line element located in \(\bm{X}\) at time \(0\) and \(\bm{l}\left(\bm{\phi}\left(\bm{X}\right), t\right)\) specifying its value at time \(t\).

Similarly, if the advection of a vector field \(\bm{s}\) satisfies
\begin{equation}
    \frac{D \bm s}{Dt}=-\left(\nabla \bm u\right)^T \bm s,
    \label{eq:surf_adv}
\end{equation}
we define \(\bm s\) as a \emph{surface element}.
The evolution of a surface element can be characterized by its bidirectional flow maps as
\begin{equation}
    \bm{s}(\bm{x}, t) = \mathcal{T}^T\,\bm{s}\left(\bm{\psi}\left(\bm{x}\right), 0\right),
    \label{eq:surf_ffm}
\end{equation}
\begin{equation}
    \bm{s}(\bm{X}, 0) = \mathcal{F}^T\,\bm{s}\left(\bm{\phi}\left(\bm{X}\right), t\right).
    \label{eq:surf_bfm}
\end{equation}

When comparing Eqn.~\eqref{eq:line_ffm} and Eqn.~\eqref{eq:line_bfm} with Eqn.~\eqref{eq:surf_ffm} and Eqn.~\eqref{eq:surf_bfm}, it becomes evident that \textit{both line and surface elements can be transported using the same underlying flow map}. The only difference is the transpose on the Jacobians for surface elements. Consequently, if an accurate flow map is constructed, it can be utilized to transport line elements, surface elements, or a combination of both, by leveraging the same geometric mapping between two time instances. This observation motivates the pursuit of the most suitable representation that, when transported using high-fidelity flow maps, could offer the highest level of numerical accuracy and stability.

For a thorough mathematical analysis of the evolution of line and surface elements in a flow field, readers might refer to the works by \citet{wu2007vorticity} and \citet{truesdell2018kinematics}, where a comprehensive discussion and valuable insights into these concepts are provided from the differential geometry perspective.

\begin{figure}[t]
\centering
\includegraphics[width=0.48\textwidth]{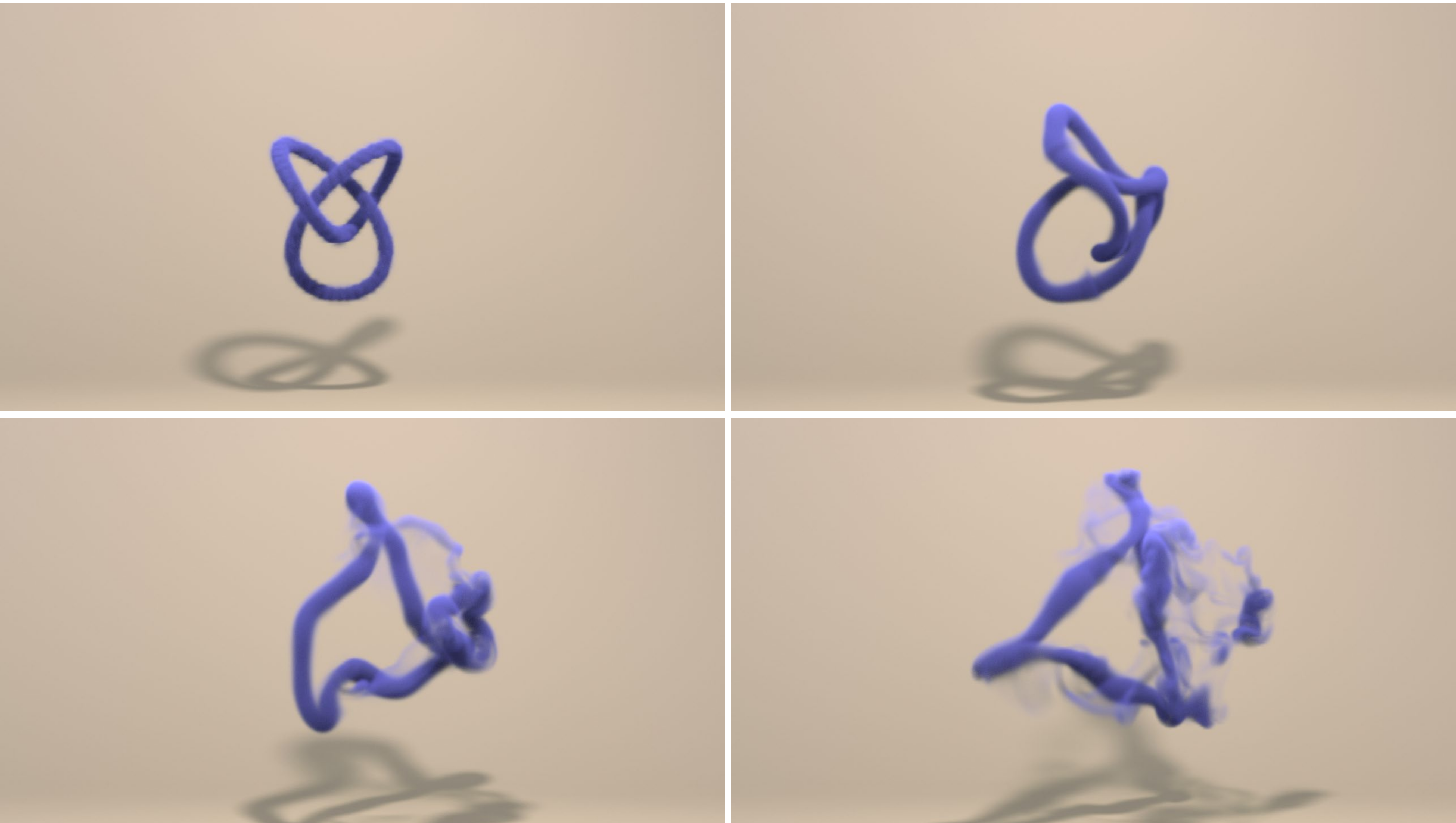}
\captionsetup{aboveskip=8pt}
\caption{The development of a vortex trefoil knot over time leads to its division into two vortex rings of varying sizes.}
\label{fig:trefoil_knot}
\vspace{-0.1in}
\end{figure}

\subsection{Inviscid Flow with Vorticity Variables}

We start with the inviscid incompressible fluid model described by the Euler equations:
\begin{equation}
    \begin{dcases}
    \frac{D \bm u }{D t} = - \frac{1}{\rho} \nabla \bm p,\\
    \nabla \cdot \bm u = 0,
    \end{dcases}
    \label{eq:ns}
\end{equation}
where $\bm u, \rho, \bm p$ represent velocity, density, and pressure respectively. The first equation specifies momentum conservation, and the second equation is the incompressibility condition.
Taking the curl of the momentum equation and substituting the incompressibility condition, we obtain the vorticity form of the Euler equation:
\begin{equation}
 \frac{D \bm \omega }{D t} =  \left( \nabla \bm u \right) \bm \omega,
 \label{eq:vor}
\end{equation}
where $\left( \nabla \bm u \right) \bm \omega$ denotes the vortex-stretching term. 
The relation between velocity and vorticity is specified by the Poisson equation
\begin{equation}
    \nabla \cdot \nabla \bm u = - \nabla \times \bm{\omega},
    \label{eq:vorticity_poisson}
\end{equation}
which at the domain boundary is subject to
\begin{equation}
    \bm u = \bm u_{wall},
    \label{eq: nonslip}
\end{equation}
and Eqn.~\eqref{eq: vort_compatibility}, where Eqn.~\eqref{eq: nonslip} is the velocity boundary condition and Eqn.~\eqref{eq: vort_compatibility} is the compatibility condition of velocity and the vorticity field in the presence of boundaries \cite{cottet2000vortex}. We refer readers to \cite{cottet2000vortex, mimeau2021review} for more details on vorticity-based incompressible fluid models.

\begin{figure}[t]
\centering
\includegraphics[width=0.48\textwidth]{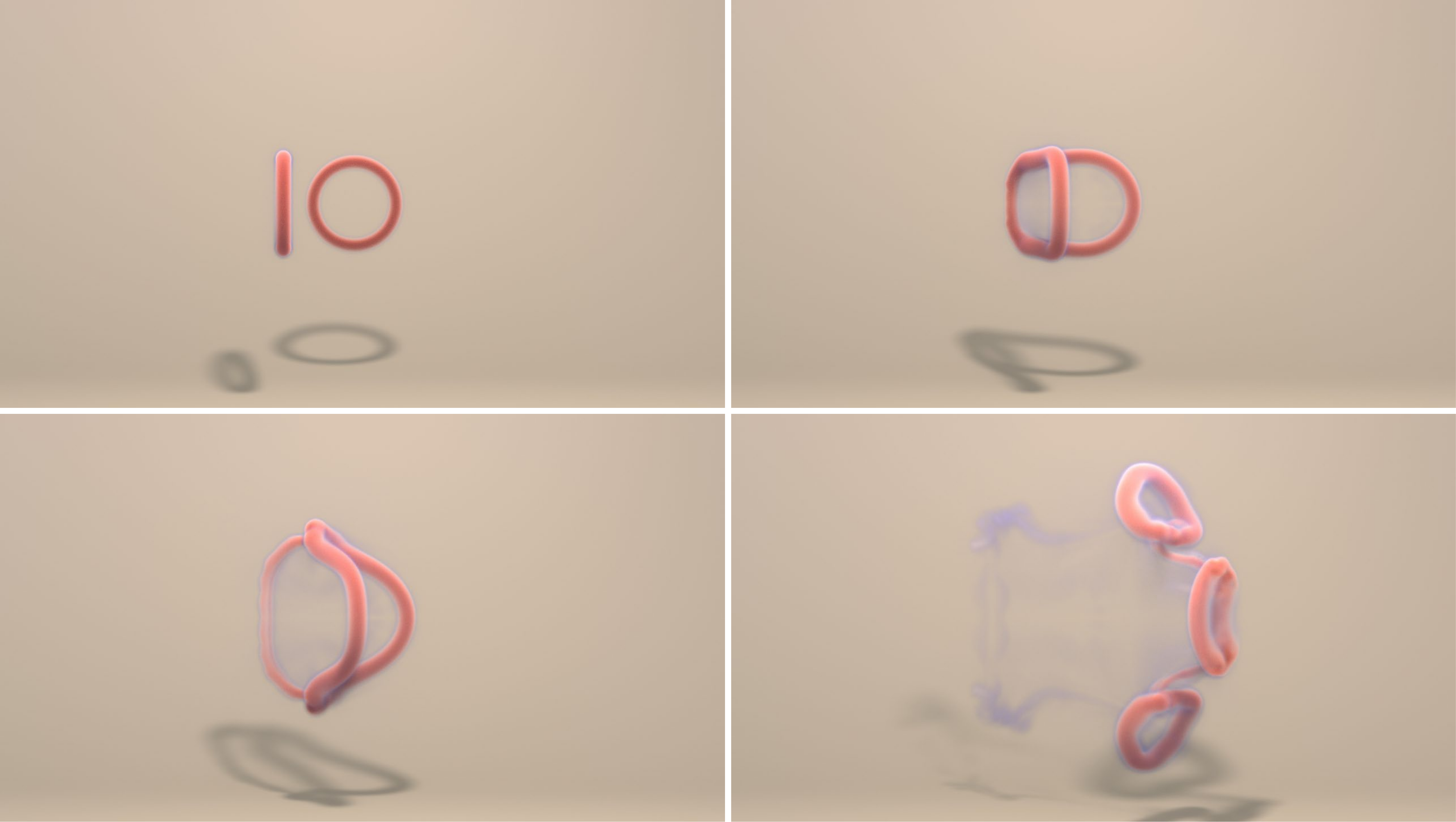}
\captionsetup{aboveskip=8pt}
\caption{The transformation of oblique vortex rings starts with two vortices colliding and merging into a single entity. It then undergoes several structural changes, eventually splitting into three distinct vortex rings.}
\label{fig:oblique_vort}
\vspace{-0.1in}
\end{figure}

\subsection{Vorticity on Flow Maps}
As evidenced in Eqn.~\eqref{eq:vor}, \emph{vorticity is a line element.} Therefore, we can leverage the flow map schemes for line elements as presented in Eqn.~\eqref{eq:line_ffm} and Eqn.~\eqref{eq:line_bfm} for its transport. Specifically, by establishing a bi-directional flow map between time \(0\) and time \(t\), we have
\begin{equation}
    \bm{\omega}\left(\bm{x}, t\right) = \mathcal{F}\,\bm{\omega}\left(\bm{\psi}\left(\bm{x}\right), 0\right),
    \label{eq:vor_ffm}
\end{equation}
\begin{equation}
    \bm{\omega}\left(\bm{X}, 0\right) = \mathcal{T}\,\bm{\omega}\left(\bm{\phi}\left(\bm{X}\right), t\right).
    \label{eq:vor_bfm}
\end{equation}
Here, in Eqn.~\eqref{eq:vor_ffm}, \(\bm{\omega}\left(\bm{x}, t\right)\) denotes the vorticity of a material point located in \(\bm{x}\) at time \(t\) and \(\bm{\omega}\left(\bm{\psi}\left(\bm{x}\right), 0\right)\) is the vorticity of the same material point at time \(0\).
Conversely, in Eqn.~\eqref{eq:vor_bfm}, \(\bm{\omega}\left(\bm{X}, 0\right)\) denotes the vorticity of a material point located in \(\bm{X}\) at time \(0\) and \(\bm{\omega}\left(\bm{\phi}\left(\bm{X}\right), t\right)\) is the vorticity of the same material point at time \(t\).
In order to establish the forward flow map Jacobian \(\mathcal{F}\) that corresponds to the trajectory from position \(\bm X\) and time \(0\) to position \(\bm x\) and time \(t\), we have to evolve \(\mathcal{F}\) in reverse.
The reverse evolution of \(\mathcal{F}\) and \(\mathcal{T}\) is governed by:
\begin{equation}
    \begin{aligned}
        \frac{D \mathcal{F}}{D t} &= \mathcal{F}\, \nabla\bm{u}, \label{eq:F_evolution_reverse} \\
        \frac{D \mathcal{T}}{D t} &= -\nabla\bm{u}\,\mathcal{T}.
    \end{aligned}
\end{equation}
For any fluid particle located at position \(\bm{x}\) at time \(t\), the result of the reverse evolution of \(\mathcal{F}\), starting from an identity map at time \(t\) with position \(\bm x\) and moving backward to time \(0\) with its original position \(\bm X\), represents the Jacobian \(\mathcal{F}\) of the forward flow map \(\bm{\phi}\) from the original position \(\bm X\) of the particle at time \(0\) to the current position \(\bm x\) at time \(t\). More details about the reverse evolution and its mathematical proof can be found in \cite{zhou2024eulerian}.

\begin{figure*}[t]
\centering
\includegraphics[width=0.994\textwidth]{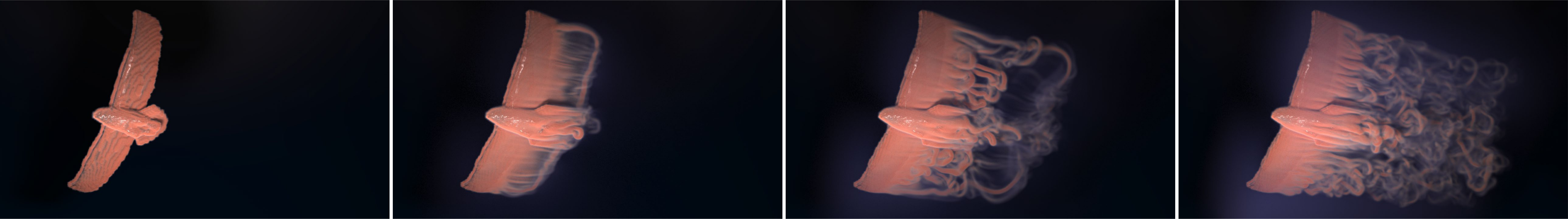}
\captionsetup{aboveskip=8pt}
\caption{The picture depicts an eagle under the behavior of gliding (256 x 256 x 128). Vortices are created at the tail and back of the wings.}
\label{fig:eagel}
\vspace{-0.1in}
\end{figure*}

\begin{figure*}[t]
\centering
\includegraphics[width=0.998\textwidth]{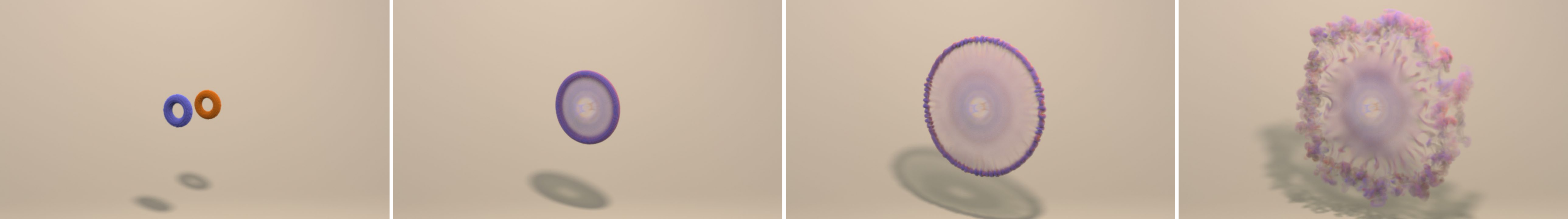}
\captionsetup{aboveskip=8pt}
\caption{Head-on vortex collision. Two vortex rings are positioned opposite to each other. They merge as one ring and expand upon collision, which ultimately leads to the ring breaking up into a series of smaller, radially arranged secondary vortices.}
\label{fig:headon_smoke}
\vspace{-0.1in}
\end{figure*}

\subsection{Relation to Impulse Variables}
To further elucidate the motivation of our design choices, we briefly describe the relationship between vorticity and impulse variables, including an illustrative example to highlight their differences in numerical performance. As indicated in the literature \cite{cortez1995impulse}, the impulse-form Euler equations write as:
\begin{equation}
\begin{dcases}
 \frac{D \bm{m}}{D t} = -\left(\nabla \bm{u}\right)^{T}\bm{m}, \\
\bm{u} = \mathcal{P}\left(\bm{m}\right),
\end{dcases}
\label{eq:imp_adv}
\end{equation}
where the first equation describes the evolution of the impulse variable \(\bm{m}\), and the second equation establishes the divergence-free condition for velocity \(\bm{u}\) by the projection operator \(\mathcal{P}\). It is obvious that \emph{impulse is a surface element} by comparing Eqn.~\eqref{eq:surf_adv} and ~\eqref{eq:imp_adv}. Consequently, the forward and backward flow maps for surface elements, as delineated in Eqn.~\eqref{eq:surf_ffm} and \eqref{eq:surf_bfm}, can be applied to determine its evolution.

\subsection{Choosing Vorticity over Impulse}

We choose to track vorticity instead of impulse on a bi-directional flow map, positing that line elements are more effective than surface elements when being numerically evolved with the same flow maps discretized on Cartesian grids. The rationale for this choice is twofold: \emph{First}, the strength of surface elements increases significantly and constantly as they evolve in a vortical flow field, which demands high precision in time integration. Specifically, as the impulse value increases, the numerical precision required by the projection solver to compute \(\bm{u}\) from \(\bm{m}\) becomes critical, making the process increasingly susceptible to the accumulation of numerical errors during impulse evolution. This issue, highlighted in several prior studies and particularly emphasized by \citet{summers1996hybrid}, is recognized as a significant barrier to using impulse-based representations to address vortex dynamics. \emph{Secondly}, the physical interpretation of vorticity directly binds to vortical structures in fluid dynamics. In particular, the distribution of vorticity is closely tied to the evolution of rotational flow motion in a fluid field, a relationship that becomes less clear when using the impulse variable to represent the same fluid motion. 

\begin{figure}[t]
\centering
\includegraphics[width=0.48\textwidth]{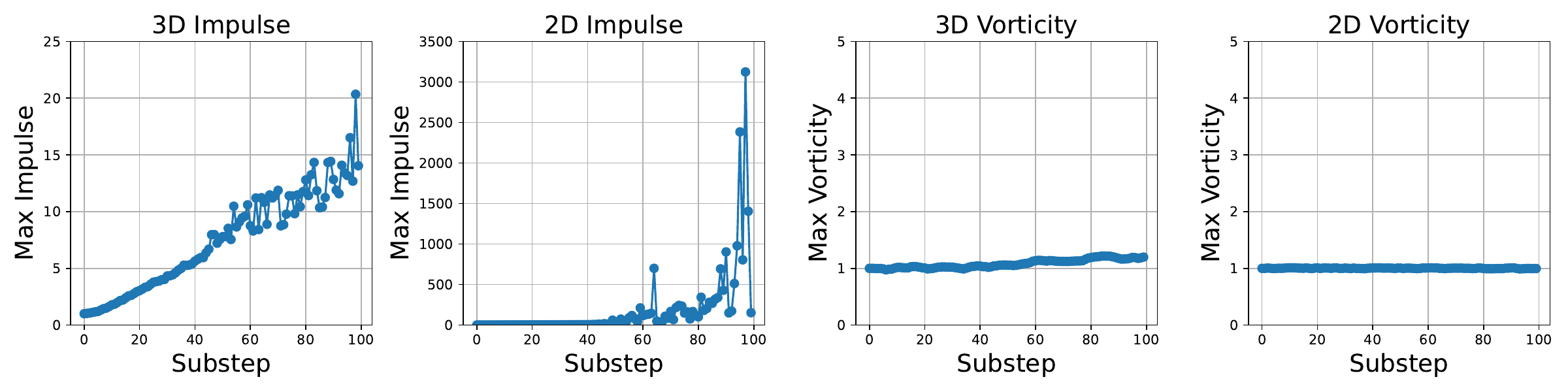}\figvspace
\caption{Motivational Experiment. The increase factor comparison of the maximum norm of impulse and vorticity in leapfrog experiments. For impulse, the maximum increase factors in 100 steps are \textbf{20.35} and \textbf{3124.27} for 3D and 2D, respectively, while the maximum increase factors for vorticity are \textbf{1.22} and \textbf{1.01}, indicating large instability induced by using impulse.}
\label{fig:impulse_enlarge}
\vspace{-0.2in}
\end{figure}

We experimentally demonstrate this insight through leapfrog experiments in 2D and 3D. As depicted in Figure~\ref{fig:impulse_enlarge}, without reinitialization, the maximum impulse increases rapidly. In contrast, the maximum vorticity remains stable throughout the simulation.

\section{Velocity Reconstruction}
As established in Section \ref{section: physical_model}, we see that vorticity $\bm \omega$ is more compatible with flow maps than impulse $\bm m$ due to its numerical stability and physical interpretability. While reconstructing $\bm u$ from $\bm m$ essentially involves a pressure projection step exactly as in conventional Eulerian methods, reconstructing $\bm u$ from $\bm \omega$ requires the solution of a vector Poisson equation as in Eqn.~\eqref{eq:vorticity_poisson}. The standard method for such reconstruction is through the streamfunction, decomposing the velocity into its vortical and potential components \cite{Ando:2015:streamfunc, elcott2007stable, zhang2015restoring, 10.1145/3592402}. However, we observe that such a decomposition is not necessary because the sole functionality of the potential component is to enforce the velocity boundary condition Eqn.~\eqref{eq: nonslip} and the compatibility condition Eqn.~\eqref{eq: vort_compatibility} near the solid boundary \cite{cottet2000vortex}. Instead, inspired by the approach with the Poincaré identity discussed in Section 4.3 of \cite{cottet2000vortex}, we directly consider the relationship between velocity and vorticity, allowing the vorticity near the solid boundaries to become unknowns to satisfy the boundary conditions. That is to say, by explicitly specifying the velocity as the Dirichlet boundary condition and implicitly enforcing the velocity-vorticity compatibility near the solid boundary, we can bypass the use of the streamfunction and the potential while still enforcing the correct boundary conditions. In this section, we propose our revised discrete formulation of the vector Poisson equation, which carefully handles the vorticity-velocity coupling near the boundary without the need for the potential component. Also, we propose a unified GPU-based solver that provides high efficiency without sacrificing physical fidelity.

\begin{figure}[t]
\centering
\includegraphics[width=0.48\textwidth]
{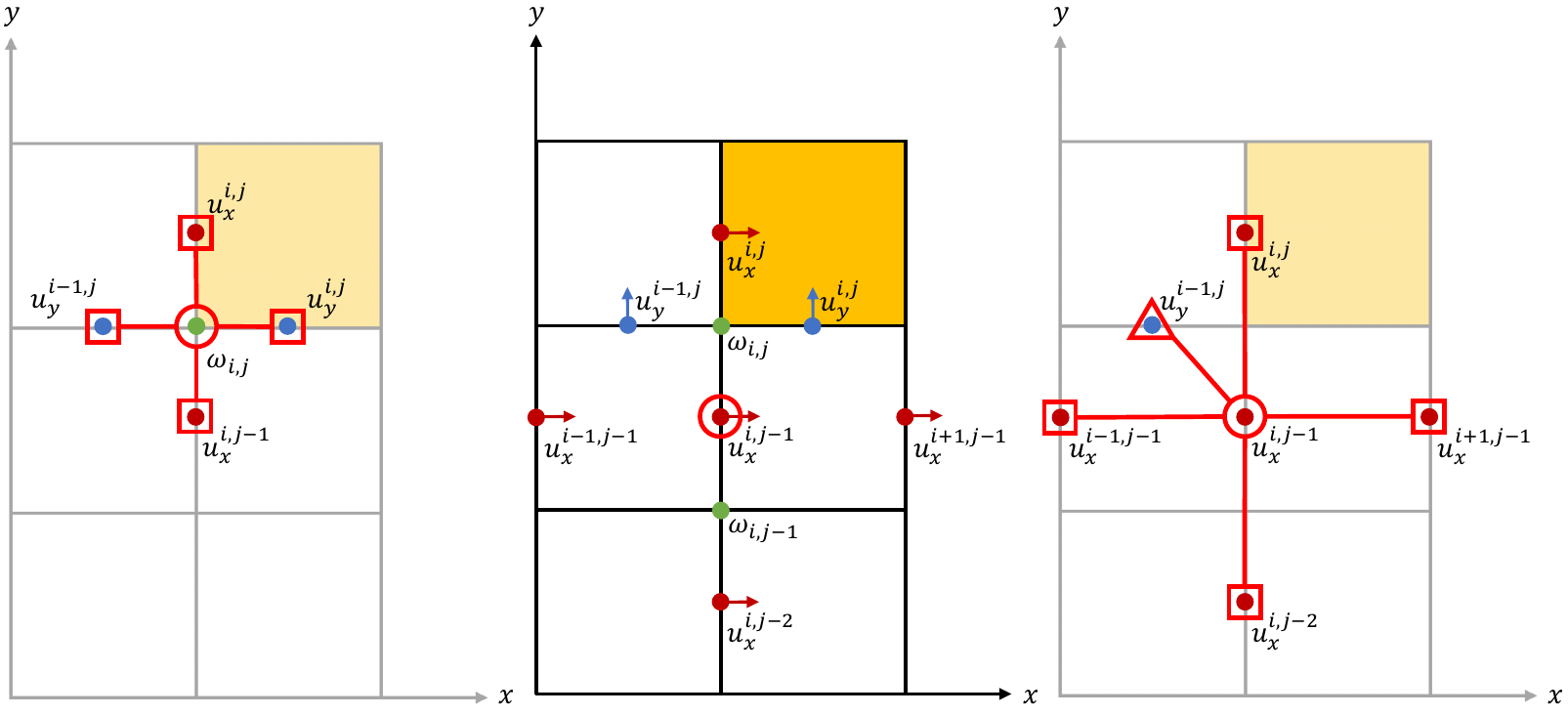}
\captionsetup{aboveskip=2pt}
\caption{Solids are represented by yellow cells. The left depicts the vorticity on the boundary, $\omega^{i, j}$ (circles), which is constrained by the four neighboring velocities (squares). The middle depicts the finite difference stencil around velocity component $u_x^{i,j-1}$, where one of the RHS variables $\omega^{i, j}$ is no longer known due to the compatibility condition. The right depicts the four neighboring variables (squares) that play a part in solving $u_x^{i, j-1}$, with the additional $u_y$ (triangles) emerging from the compatibility condition.}
\label{fig:solid_boundary}
\vspace{-0.17in}
\end{figure}

\subsection{Solid Boundary Condition}
\label{sec:sec4.1}

In this section, we concretely illustrate the challenge in handling the boundary conditions in formulating the Poisson equation, by considering the 2D case as depicted in Figure~\ref{fig:solid_boundary}. By discretizing Eqn.  \eqref{eq:vorticity_poisson} using finite difference on a staggered marker-and-cell (MAC) grid as shown in Figure~\ref{fig:solid_boundary}, we can write it out as a system of scalar equations on staggered locations. We use the equation at \(u_x^{i, j-1}\) as an example:

 \begin{equation}
 \begin{aligned}
 \label{eq:2D_discretization}
4 u_{x}^{i, j-1} - u_{x}^{i+1, j-1} - u_{x}^{i, j-2} - u_{x}^{i-1, j-1} - u_{x}^{i, j}
&= (\omega^{i, j} - \omega^{i, j-1}) \cdot \Delta x,
\end{aligned}
 \end{equation}
where \(u_x^{i, j-1}\) represents the horizontal velocity on the face $(i, j-1)$. The system can be solved if the RHS contains only known variables.

In cases where both $\omega^{i, j}$ and $\omega^{i, j-1}$ are in the fluid's interior, they are indeed known variables that are prescribed by Eqn. \eqref{eq:vor_ffm}.
However, as illustrated in Figure~\ref{fig:solid_boundary} (left), this ceases to be the case near the boundary, since the velocity boundary condition Eqn.~\eqref{eq: nonslip} propagates to the vorticity $\bm \omega$ through the compatibility condition  Eqn.~\eqref{eq: vort_compatibility}. The compatibility requires that the velocity and the vorticity field need to satisfy Eqn.~\eqref{eq: vort_compatibility} everywhere in the simulation domain. Specifically, consider Eqn.~\eqref{eq:2D_discretization} for variable $u_x^{i,j-1}$ in the middle of Figure~\ref{fig:solid_boundary}, which involves a RHS variable $\omega^{i, j}$ that is constrained to its neighboring velocities $u_y^{i-1, j}$, $u_y^{i, j}$, $u_x^{i, j-1}$, $u_x^{i, j}$ by the compatibility condition:
\begin{equation}
\label{eq:exp_vort_by_vel}
    \omega^{i, j} = \frac{(u_{y}^{i, j} - u_{y}^{i-1, j}) - (u_{x}^{i, j} - u_{x}^{i, j-1})}{\Delta x}.
\end{equation}
Since both $u_y^{i, j}$ and $u_{x}^{i, j}$ are on solid faces, they are dictated by the wall velocities $u_{y,wall}^{i, j}$ and $u_{x, wall}^{i, j}$. Consequently, $\omega^{i, j}$ becomes an unknown variable dependent on $u_{x}^{i, j-1}$ and $u_{y}^{i-1, j}$.

\begin{figure}[t]
 \centering
 \includegraphics[width=0.48\textwidth]{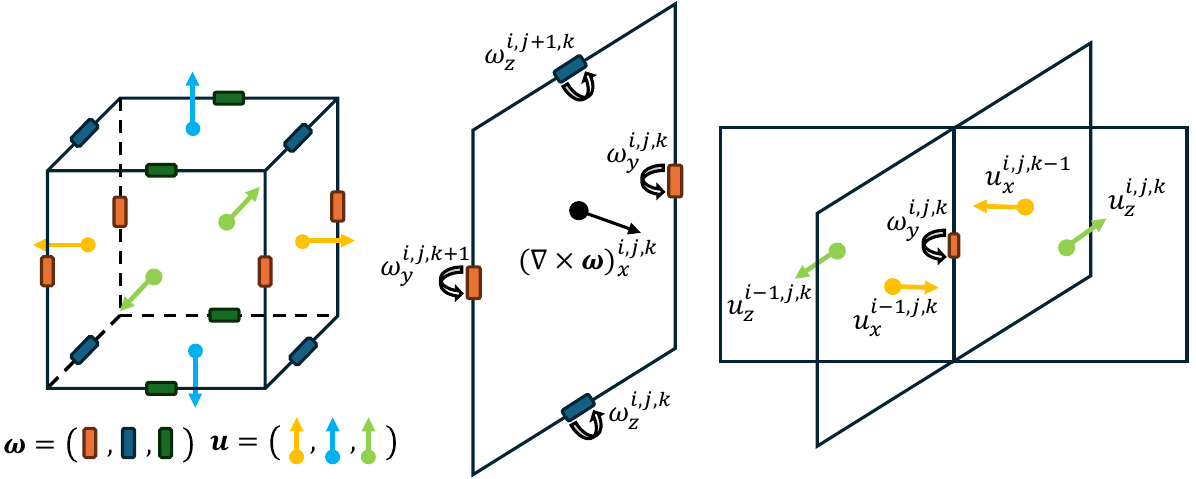}
 \captionsetup{aboveskip=2pt}
 \figvspace
 \caption{Illustration of our discretization scheme. The curl operator maps edges to faces (middle) or faces to edges (right).}
 \label{fig:disc}
 \vspace{-0.1in}
\end{figure}

\begin{figure}[t]
\centering
\includegraphics[width=0.48\textwidth]{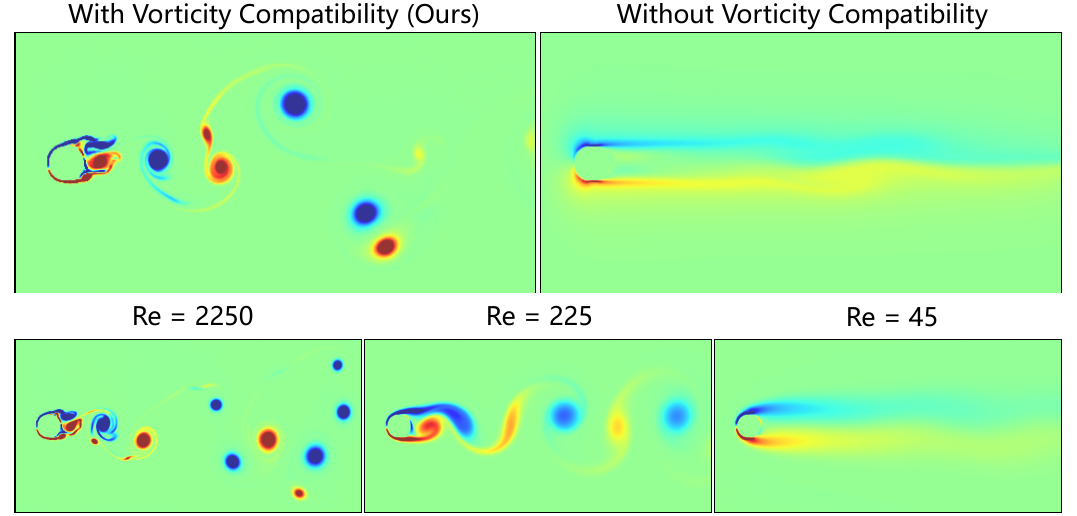}
\captionsetup{aboveskip=4pt}
\caption{The top row shows an ablation study for the velocity-vorticity compatibility, which is the comparison of the $150^\text{th}$ frame of the 2D von Kármán vortex street example with a high Reynolds number (Re) = 2250. The bottom row shows that our method accurately models the 2D von Kármán vortex street with varying Re.}
\label{fig:kvt_solid_boundary_comp}
\vspace{-0.12in}
\end{figure}

\subsection{Revised Poisson Equation}
To this end, we need to eliminate the unknown variables on the RHS of Eqn.~\eqref{eq:2D_discretization}. By substituting \(\omega^{i, j}\) on the RHS, we obtain the revised equation for $u_{x}^{i, j-1}$:
\begin{equation}
\begin{aligned}
\label{eq:voxelized_solid_vort_in_solid_case_1}
3 u_{x}^{i, j-1} - u_{x}^{i+1, j-1} - u_{x}^{i, j-2} - u_{x}^{i-1, j-1} + u_{y}^{i-1, j}\\
    =  {u_{y, wall}^{i, j}}  - \omega^{i, j-1} \cdot \Delta x \text{.}
\end{aligned}
\end{equation}

It can be observed from the fact that $u_{x}^{i, j-1}$ is now dependent on $u_{y}^{i-1, j}$, that such a revised system of equations will be coupled across spatial dimensions, i.e., it can no longer be solved using a generic scalar Poisson solver by treating each dimension separately. For the general case in 3D, \(u_x\) has six neighboring velocities, with two on each axis. The y and z-axis neighbors differ from the x-axis neighbors due to an intervening vorticity variable. If this variable is not within the fluid (e.g., on a solid boundary), we adjust the LHS by subtracting one from the coefficient of \(u_x\), removing the corresponding velocity, and adding or subtracting unknowns from other axes within the fluid depending on their relative positions, as shown by an analogous 2D example in Figure~\ref{fig:solid_boundary}. Then, we adjust the RHS by removing such intervening vorticities and moving the known velocities on the walls to the RHS. More details and algorithms are discussed in the Appendix Section~\ref{sec:app_multigrid}.

\subsection{Coupled Poisson Solver}
To solve this revised equation system efficiently and accurately, we develop a GPU-based matrix-free Multi-Grid Preconditioned Conjugate Gradient (MGPCG) solver, whose implementation details are given in the supplementary material. Our solver is a geometric multi-grid solver implemented in Taichi \cite{hu2019taichi}, with solid boundary conditions being enforced at coarser levels with coarsened solids. The restriction, prolongation, and smoothing are implemented following \cite{mcadams2010parallel}.
On one hand, as seen in Figure~\ref{fig:kvt_solid_boundary_comp} (top), our method is able to significantly improve the simulation realism with its physically correct treatment of vorticity-velocity coupling near the boundary. On the other hand, as shown in Table~\ref{tab: couple_table}, by leveraging the matrix-free nature of our solver, we are able to conveniently, without incurring a hindering memory overhead, harness the parallel computation of modern GPUs to process multiple dimensions concurrently.

\section{Numerical Algorithm}
Leveraging our novel vorticity-to-velocity method as its core, we build a highly accurate simulation method. As shown in Algorithm~\ref{alg:main}, our simulation time step is completed in the following stages:
\begin{enumerate}
    \item Compute forward and backward flow maps (steps 2-7);
    \item Compute vorticity advection using flow maps (step 8);
    \item Compute velocity from vorticity (step 9).
\end{enumerate}

\subsection{Discretization}

As shown in Figure~\ref{fig:disc}, \(\bm u\) and \(\bm \omega\) are stored on faces and edges, in line with their physical semantics as line and surface elements. Differential operators such as curl and divergence can be defined based on Stokes's Theorem. We refer to \cite{hirani2003discrete} for details.

\begin{algorithm}
\caption{Time Integration}
\label{alg:main}
\begin{flushleft}
        \textbf{Initialize:} $\bm{u}$, $\bm{w}$ to initial velocity and vorticity; 
\end{flushleft}
\begin{algorithmic}[1]
\While{simulating}
\State \textbf{Reinitialize} every $n$ steps;
\State Compute $\Delta t$ with $\bm{u}$ and CFL number;
\State $\bm{u}_\text{mid} \gets$ \textbf{MidPoint}$(\bm u, \bm{\omega}, \Delta t)$;
\State Store $\bm u_\text{mid}$ in the buffer $\mathcal{V}$ and $\Delta t$ in $\mathcal{S}$;
\State $\bm{\psi}, \mathcal{F} \gets$ \textbf{Backtrace}$(\mathcal{V}, \mathcal{S})$;
\State $\bm{\phi}, \mathcal{T} \gets$ \textbf{March}$(\bm u_\text{mid}, \Delta t, \bm{\phi}, \mathcal{T})$;
\State $\hat{\bm{\omega}} \gets$ \textbf{BFECC}$(\bm{\omega}_\text{init}, \bm{\psi}, \mathcal{F}, \bm{\phi}, \mathcal{T})$;
\State $\bm{u} \gets \Delta^{-1} (-\nabla \times \hat{\bm{\omega}})$.
\EndWhile
\end{algorithmic}
\end{algorithm}
\begin{algorithm}
\caption{Vorticity Flow Map Computation for Step \(j\)}
\label{alg:flow_map_comp_w}
\begin{flushleft}
        \textbf{Input:} $j$, $\mathcal{V}$, $\mathcal{S}$, $\bm{\phi}$, $\mathcal{T}$.\\
        \textbf{Output:} $\bm{\psi}$, $\mathcal{F}$, $\bm{\phi}$, $\mathcal{T}$.
\end{flushleft}
\begin{algorithmic}[1]

\State Estimate midpoint velocity $\bm{u}_\text{mid}$ according to Algorithm \ref{alg:midpoint};
\State $\bm{\psi} \gets \textbf{id}_\Omega$, $\mathcal{F} \gets \bm{I}$;
\For {$l$ in j \dots 0}
\State Fetch $\bm{u}_{\text{mid}, l}$ and $\Delta t_l$ from the buffer $\mathcal{V}$ and $\mathcal{S}$;
\State March $\bm{\psi}, \mathcal{F}$ with Algorithm~\ref{alg:RK4_w}, using $\bm{u}_{\text{mid}, l}$ and $-\Delta t_l$;
\EndFor{}
\State March $\bm{\phi}, \mathcal{T}$ with Algorithm~\ref{alg:RK4_w}, using $\bm{u}_\text{mid}$ and $\Delta t_j$.
\end{algorithmic}
\end{algorithm}

\begin{algorithm}
\caption{Vorticity Joint RK4 for $\bm{\psi}$ and $\mathcal{F}$}
\label{alg:RK4_w}
\begin{flushleft}
        \textbf{Input:} $\bm{u}$, $\bm{\psi}$, $\mathcal{F}$, $\Delta t$.~\\
        \textbf{Output:} $\bm{\psi}_\text{next}$, $\mathcal{F}_\text{next}$.
\end{flushleft}
\begin{algorithmic}[1]
\State $(\bm{u}_1, \grad \bm{u}\vert_1) \gets \textbf{Interpolate}(\bm{u}, \bm{\psi})$;
\State $\frac{\partial \mathcal{F}}{\partial t}\vert_1 \gets \mathcal{F}\, \grad \bm{u}\vert_1$;
\State $\bm{\psi}_1 \gets \bm{\psi} + 0.5 \cdot \Delta t \cdot \bm{u}_1$;
\State $\mathcal{F}_1 \gets \mathcal{F} - 0.5 \cdot \Delta t \cdot \frac{\partial \mathcal{F}}{\partial t}\vert_1$;
\State $(\bm{u}_2, \grad \bm{u}\vert_2)\gets \textbf{Interpolate}(\bm{u}, \bm{\psi}_1)$;
\State $\frac{\partial \mathcal{F}}{\partial t}\vert_2 \gets \mathcal{F}_1 \grad \bm{u}\vert_2$;
\State $\bm{\psi}_2 \gets \bm{\psi} + 0.5 \cdot \Delta t \cdot \bm{u}_2$;
\State $\mathcal{F}_2 \gets \mathcal{F} - 0.5 \cdot \Delta t \cdot \frac{\partial \mathcal{F}}{\partial t}\vert_2$;
\State $(\bm{u}_3, \grad \bm{u}\vert_3)\gets \textbf{Interpolate}(\bm{u}, \bm{\psi}_2)$;
\State $\frac{\partial \mathcal{F}}{\partial t}\vert_3 \gets \mathcal{F}_2 \grad \bm{u}\vert_3$;
\State $\bm{\psi}_3 \gets \bm{\psi} + \Delta t \cdot \bm{u}_3$;
\State $\mathcal{F}_3 \gets \mathcal{F} - \Delta t \cdot \frac{\partial \mathcal{F}}{\partial t}\vert_3$;
\State $(\bm{u}_4, \grad \bm{u}\vert_4)\gets \textbf{Interpolate}(\bm{u}, \bm{\psi}_3)$;
\State $\frac{\partial \mathcal{F}}{\partial t}\vert_4 \gets \mathcal{F}_3 \grad \bm{u}\vert_4$;
\State $\bm{\psi}_\text{next} \gets \bm{\psi} + \Delta t \cdot \frac{1}{6}\cdot (\bm{u}_1 + 2 \cdot \bm{u}_2 + 2 \cdot \bm{u}_3 + \bm{u}_4)$;
\State $\mathcal{F}_\text{next} \gets \mathcal{F} - \Delta t \cdot  \frac{1}{6} \cdot (\frac{\partial \mathcal{F}}{\partial t}\vert_1 
 + 2 \cdot \frac{\partial \mathcal{F}}{\partial t}\vert_2 + 2 \cdot\frac{\partial \mathcal{F}}{\partial t}\vert_3 + \frac{\partial \mathcal{F}}{\partial t}\vert_4)$.
\end{algorithmic}
\end{algorithm}

\begin{algorithm}
\caption{Vorticity Midpoint Method}
\label{alg:midpoint}
\begin{flushleft}
        \textbf{Input:} $\bm{u}$, $\bm{\omega}$, $\Delta t$.~\\
        \textbf{Output:} $\bm{u}_\text{mid}$.
\end{flushleft}
\begin{algorithmic}[1]
\State Reset $\bm{\psi}, \mathcal{F}$ to identity;
\State March $\bm{\psi}, \mathcal{F}$ with Algorithm~\ref{alg:RK4_w}, using $\bm{u}$ and $-0.5 \Delta t$;
\State $\bm{\omega}_\text{mid} \gets \mathcal{F}\bm{\omega}(\bm{\psi})$;
\State $\bm{u}_\text{mid} \gets \textbf{Poisson}({\bm{\omega}}_\text{mid})$.
\end{algorithmic}
\end{algorithm}

\begin{algorithm}
\caption{Error-compensated Vorticity Pullback}
\label{alg:bfecc}
\begin{flushleft}
        \textbf{Input:} $\bm{\omega}_{init}$, $\bm{\psi}$, $\mathcal{T}$, $\bm{\phi}$, $\mathcal{F}$.~\\
        \textbf{Output:} $\bm{u}$.
\end{flushleft}
\begin{algorithmic}[1]
\State $\bar{\bm{\omega}} \gets \mathcal{F}\bm{\omega}_{init}(\bm{\psi})$;
\State $\bar{\bm{\omega}}_{init} \gets \mathcal{T}\bar{\bm{\omega}}(\bm{\phi})$;
\State $\bm{e} \gets 0.5 \cdot (\bar{\bm{\omega}}_{init} - \bm{\omega}_{init})$;
\State $\bar{\bm{e}} \gets \mathcal{F}\bm{e}(\bm{\psi})$;
\State $\hat{\bm{\omega}} \gets \bm{\bar{\omega}} - \bar{\bm{e}}$;
\State $\bm{\omega} \gets \textbf{Clamp}(\hat{\bm{\omega}})$;
\State $\bm{u} \gets \textbf{Poisson}(\bm{\omega})$.
\end{algorithmic}
\end{algorithm}

\subsection{Time Integration}
\paragraph{Bi-directional March}
For each step, we first construct flow map quantities using all previously stored mid-point velocity fields. In this process, \( \bm{\phi} \) and \( \mathcal{T} \) are evolved forward by the Runge-Kutta method. We adopt the bi-directional march from \cite{deng2023fluid}. Consequently, \( \bm{\psi} \) and \( \mathcal{F} \) are evolved backward in the same manner as \( \bm{\phi} \) and \( \mathcal{T} \), with time reversed. Note that, unlike in \cite{deng2023fluid} where $\bm{\psi}$ and $\mathcal{T}$ are evolved together, we now need to evolve $\mathcal{F}$ in reverse with \(\bm \psi\), due to the difference in Eqn.~\eqref{eq:line_ffm} and Eqn.~\eqref{eq:surf_ffm}. The detailed algorithm is outlined in Algorithm~\ref{alg:flow_map_comp_w} and Algorithm~\ref{alg:RK4_w}.
\begin{figure*}[t]
\centering
\includegraphics[width=0.995\textwidth]{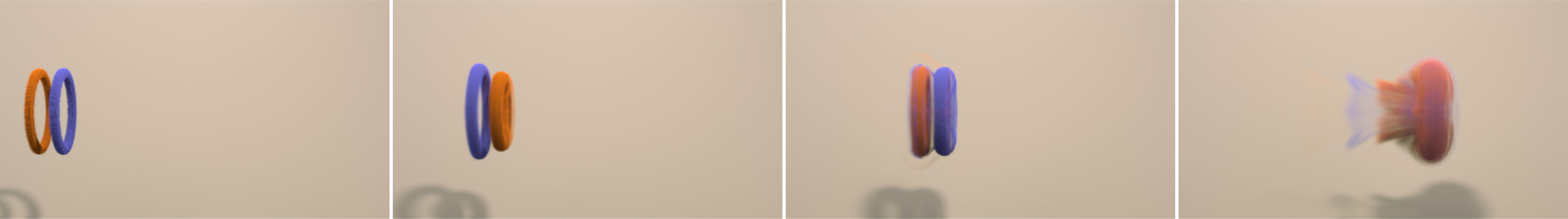}
\figvspace
\caption{Leapfrog vortices in 3D. The two rings remain separated after the $4^\text{th}$ leap.}
\label{fig:3D_leapfrog}
\end{figure*}

\begin{figure}[t]
\centering
\includegraphics[width=0.48\textwidth]{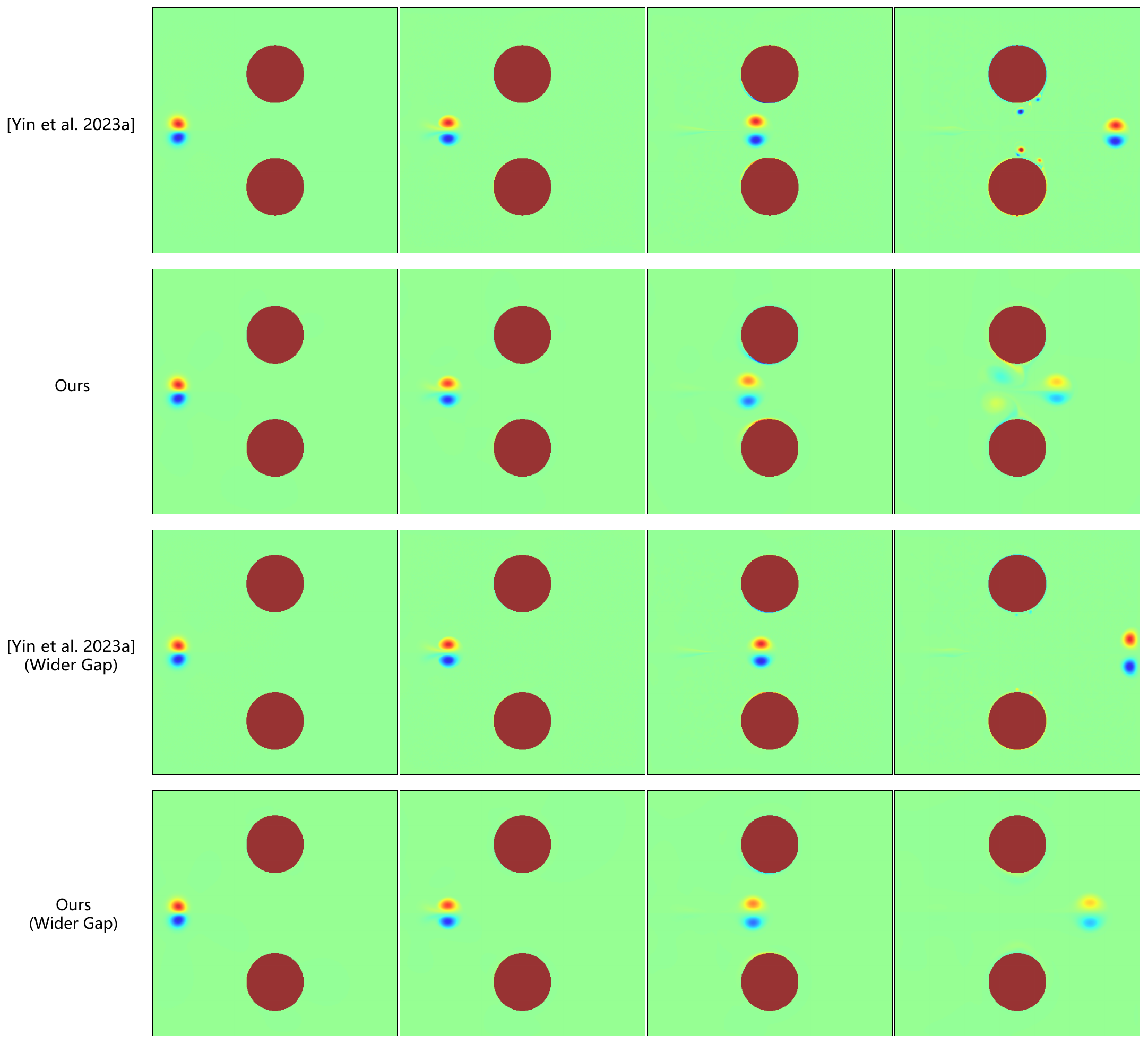}
\captionsetup{aboveskip=8pt}
\caption{A vortex pair successfully passes through a gap between two disks, demonstrating that by implicitly modeling the harmonic component, our result is consistent with Fluid Cohomology \cite{10.1145/3592402}, where the harmonic component is explicitly modeled.}
\label{fig:harmonic_4}
\end{figure}

\paragraph{Midpoint Method}
For time integration, We adopt the midpoint method, following \citet{nabizadeh2022covector}. To obtain \(\bm{\psi}\) and \(\mathcal{F}\) for advecting the vorticity, we first set \(\bm{\psi}\) and \(\mathcal{F}\) to its original position and identity map respectively. Then \(\bm{\psi}\) and \(\mathcal{F}\) are evolved backward together by a half step. Finally, we can utilize Eqn.~\eqref{eq:vor_ffm} to compute the midpoint vorticity. The detailed algorithm is outlined in Algorithm~\ref{alg:midpoint}.


\paragraph{Vorticity Advection using Flow Maps}
Leveraging the highly accurate bi-directional flow map, we adopt BFECC \cite{kim2005flowfixer} to perform advection in both directions.
We begin by using the backward flow map \(\bm{\psi}\) to advect the initial vorticity field to the current time step, represented as
$
    \bm \omega \left(\bm x, t\right)=\mathcal{F}\,\bm{\omega} \left(\bm{\psi} \left(\bm x\right), 0\right).
    \label{eq:imp_map}
$
Next, the forward flow map \(\bm{\phi}\) is applied to advect \(\bm{\omega} \left(\bm{x}, t\right)\) back to the initial time step, given by
$
    \bar{\bm \omega} \left(\bm X, 0\right)=\mathcal{T}\,\bm \omega \left(\bm{\phi} \left(\bm X\right), t\right).
$
Then we compare the error between $\bar{\bm \omega} \left(\bm X, 0\right)$ and ${\bm \omega} \left(\bm X, 0\right)$ for computing the BFECC error. The detailed algorithm is outlined in Algorithm~\ref{alg:bfecc}.

\subsection{Implementational Details}
\paragraph{Reinitialization.}
During the simulation, the flow map quantities may undergo significant distortion, necessitating reinitialization every $n$ steps, where $n$ is as a parameter which may be tuned for different simulation scenarios. At every reinitialization step, the initial vorticity \(\bm \omega_0\) is set to the curl of the current velocity. The forward map \(\bm{\phi}\) and the backward Jacobian \(\mathcal{T}\) are set to the grid positions and identity map.

\paragraph{External forces and viscosity.}
We adopt the accumulation buffer from \cite{qu2019efficient} for external forces and viscosity. In each step, we compute the current vorticity change and add this change back to the buffer at its undeformed position with forward map $\bm{\phi}$.

\begin{figure}
\centering
\includegraphics[width=0.48\textwidth]{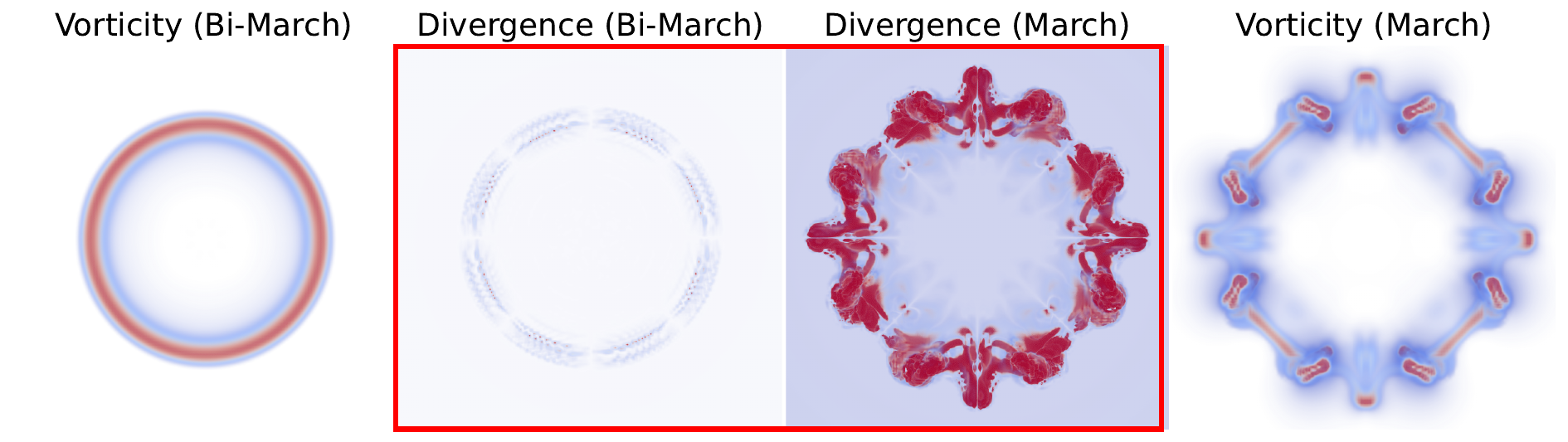}
\caption{\textbf{Vorticity Divergence Test}. Here we show an ablation study for the bi-directional marching applied for the Eulerian vortex method, which shows the comparison of the $50^\text{th}$ frame of the head-on vortex collision for the divergence of vorticity, which is theoretically \textbf{zero} as it is the curl of velocity. Without a bi-directional march (right), the highest divergence is \textbf{45538.9922}. With bi-directional march (left), the highest divergence significantly reduces to \textbf{559.9824}, a reduction by nearly \textbf{80} times. A clear contrast can be seen between the two images highlighted with red boxes.}
\label{fig:vor_div}
\end{figure}

\begin{figure}
\centering
\includegraphics[width=0.48\textwidth]{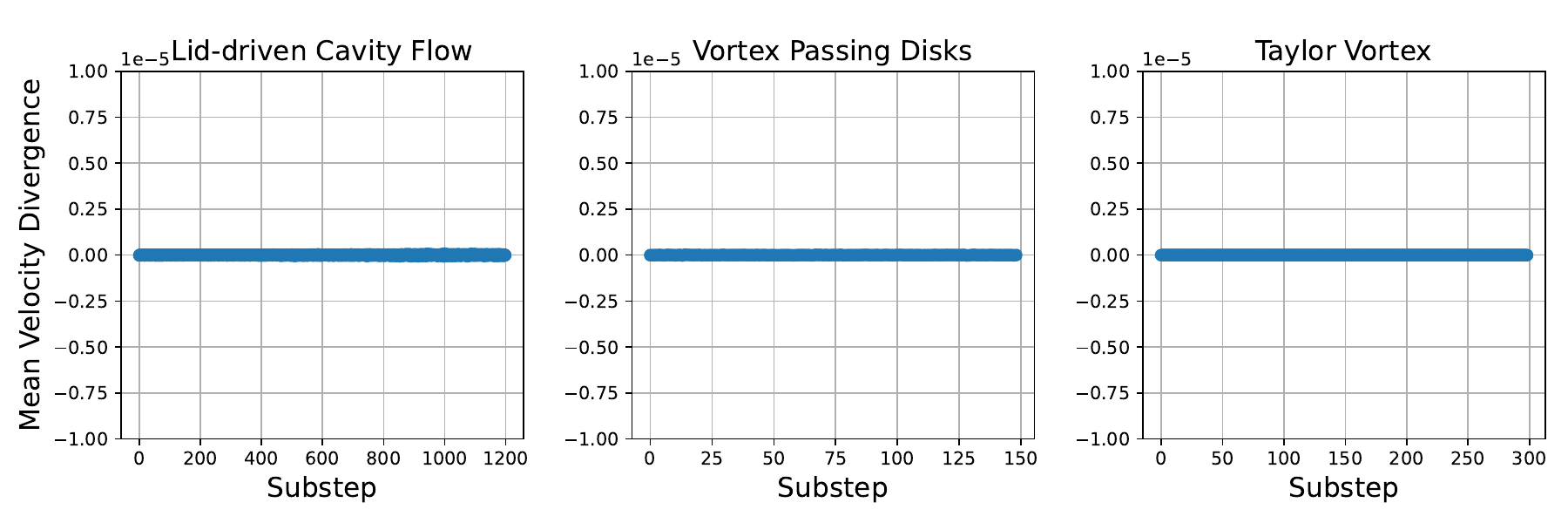}
\caption{\textbf{Velocity Divergence Test}. Using three examples, we verify that our reconstructed velocity field is divergence free.}
\label{fig:vel_div}
\end{figure}

\section{Evaluation}

In this section, we will first verify the correctness of our revised Poisson scheme and boundary treatment; we will then provide the performance gains of our coupled approach and the performance measurements; next, we will compare our method to standard 2D and 3D benchmarks; finally, we will exhibit our simulation results with complex 3D scenarios with moving solid objects. All experimental settings are included in the Appendix Table~\ref{tab: examples_table}. Compared to existing vortex methods, our approach demonstrates significantly improved capabilities in capturing physical phenomena and preserving vortical details. All performance analysis are conducted on a laptop equipped with a 13th Gen Intel(R) Core(TM) i9-13900HX processor (2.20 GHz), 32 GB of RAM, and an NVIDIA GeForce RTX 4080 Laptop GPU. We use Taichi \cite{hu2019taichi} version 1.6.0 as our parallel programming tool.

\subsection{Solid Boundary Experiments and Divergence Test}

\vspace{0.1cm}
\noindent
\textbf{2D von Kármán vortex street.}
Figure~\ref{fig:kvt_solid_boundary_comp} (bottom) shows fluid inflow at \(0.16 \, \text{m/s}\) interacting with a disk of \(0.141 \, \text{m}\) diameter, resulting in three vortex shedding patterns under varying viscosities. Our method accurately replicates diverse shedding behaviors across different Reynolds numbers (Re), in line with physical experiments \cite{blevins1990flow}. Simulations exhibit turbulent mixing at high Re (\(2250\)), a periodic vortex street at moderate Re (\(225\)), and a laminar wake without vortices at low Re (\(45\)). This aligns well with the observed behavior across Re = 60--5000~\cite{wu2003streamline}.

\vspace{0.1cm}
\noindent
\textbf{Velocity-vorticity compatibility.}
Figure~\ref{fig:kvt_solid_boundary_comp} (top) shows a comparison of the $150^\text{th}$ frame of the 2D von Kármán vortex street example with a high Re (2250). The right shows the boundary treatment that only enforces the velocity boundary condition. Without the velocity-vorticity compatibility, the system fails to generate vortices.

\vspace{0.1cm}
\noindent
\textbf{Harmonic component.}
Figure~\ref{fig:harmonic_4} shows the vorticity visualization of a vortex pair passing through two disks in 2D, along with a comparison with Fluid Cohomology \cite{10.1145/3592402}, which explicitly models the harmonic component. As described in \cite{10.1145/3592402}, the expected behavior for the vortex pair of a vortex method that correctly handles the harmonic component is to pass through the two disks. Ours can pass through the two disks as expected, further verifying the correctness of our method, though with a bit more dissipation compared to \cite{10.1145/3592402}.

\begin{wrapfigure}{r}{0.1\textwidth}
\centering
\includegraphics[width=0.1\textwidth]{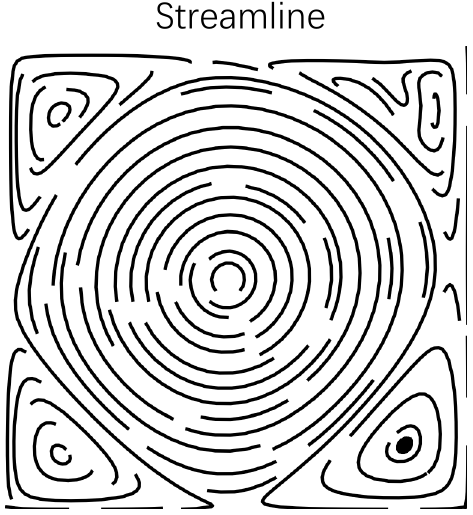}
\caption{Cavity flow}
\label{fig:cavity_comp}
\vspace{-0.1in}
\end{wrapfigure}

\vspace{0.1cm}
\noindent
\textbf{Lid-driven cavity flow.}
Figure~\ref{fig:cavity_comp} shows the streamline of the 2D Lid-driven cavity flow where the velocity of the top boundary is fixed at 1m/s horizontally to the right, while the other three boundaries are static solid walls. We tested our method with Re = 5000, it generates three vortices at the left top, left bottom and right bottom corner, which is consistent with the ground truth shown in \cite{ghia1982high}.

\vspace{0.1cm}
\noindent
\textbf{Velocity Divergence.}
Figure~\ref{fig:vel_div} shows the mean divergence of three examples. Even though we do not use the vorticity-streamfunction formulation, the divergence remains near zero.

\vspace{0.1cm}
\noindent
\textbf{Vorticity Divergence.}
We demonstrate the effectiveness of bi-directional marching when applied to our method. A significant challenge associated with a traditional Eulerian vortex method is the accumulation of numerical errors over time, leading to a vorticity field that is not divergence-free. As illustrated in Figure~\ref{fig:vor_div}, adopting the bi-directional march leads to a nearly \textbf{80}$\times$ decrease in error.

\begin{figure}[t]
\centering
\includegraphics[width=0.48\textwidth]{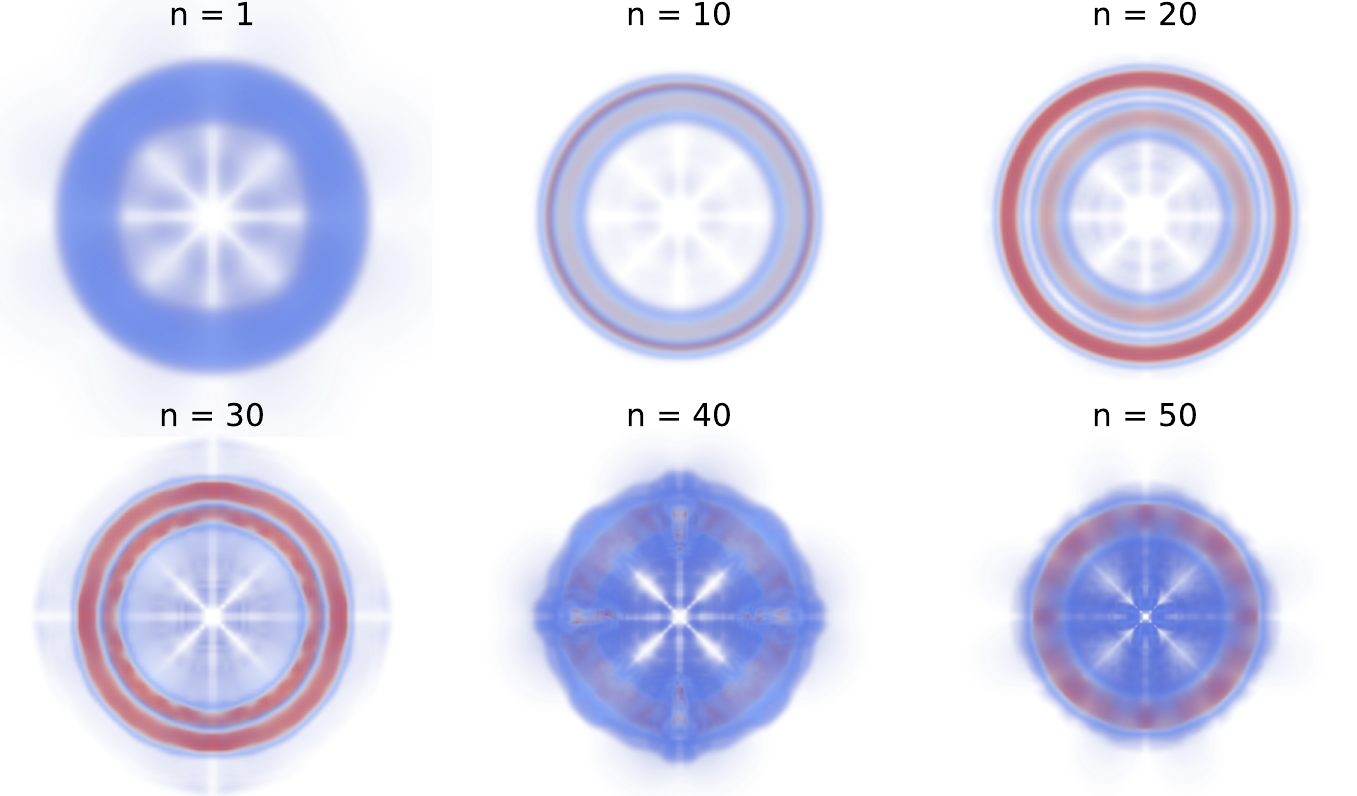}
\captionsetup{aboveskip=8pt}
\figvspace
\caption{Reinitialization sensitivity experiment. The $150^\text{th}$ frame of the 3D leapfrog with different reinitialization step $n$ is shown.}
\label{fig:n_sense}
\vspace{-0.1in}
\end{figure}

\begin{figure}[t]
\centering
\includegraphics[width=0.48\textwidth]{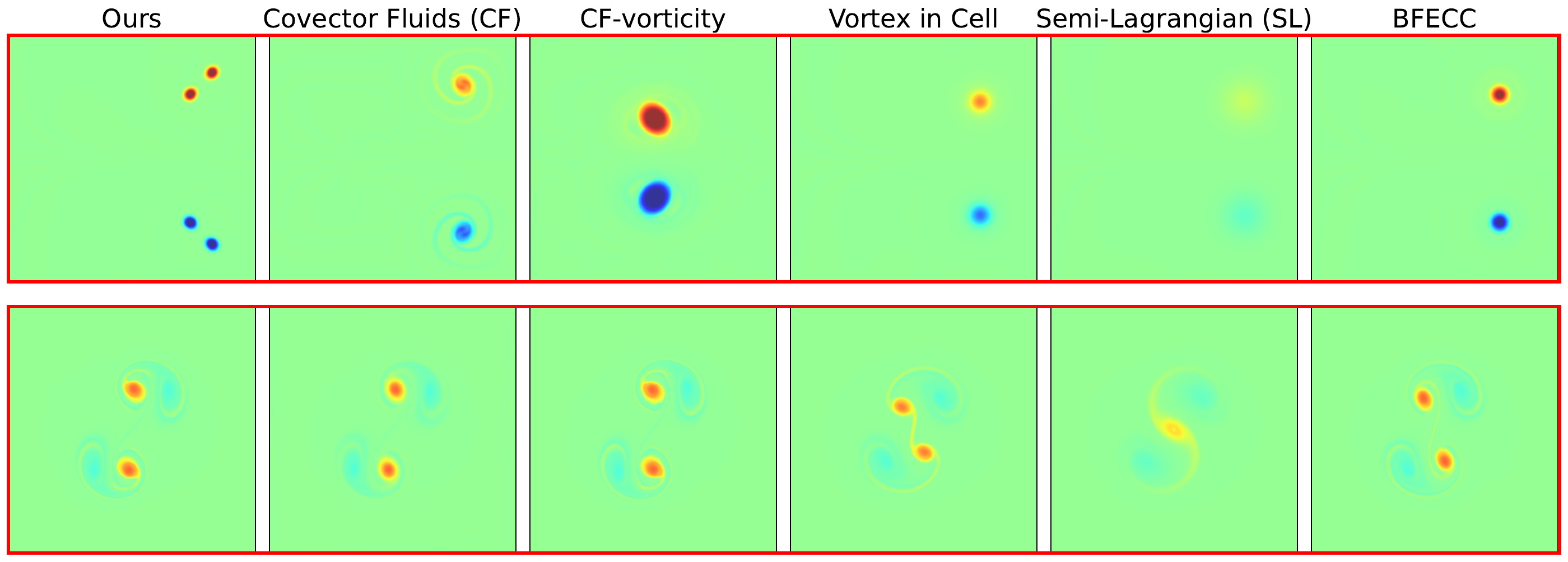}
\captionsetup{aboveskip=8pt}
\caption{Comparisons of 2D leapfrog vortices (top) and Taylor vortex (bottom).
The columns from left to right depict ours, Covector Fluids (CF) \cite{nabizadeh2022covector}, the vorticity version of CF (traditional flow map advection), Vortex-in-Cell, Semi-Lagrangian (SL) advection, Back and Forth Error Compensation and Correction (BFECC) respectively (SL and BFECC only refer to the advection techniques used, so these are still vortex methods. A normal velocity-pressure BFECC comparison is shown in Figure \ref{fig:3D_leapfrog_compare}).}
\label{fig:2D_comparison}
\end{figure}

\begin{figure}[t]
\centering
\includegraphics[width=0.48\textwidth]{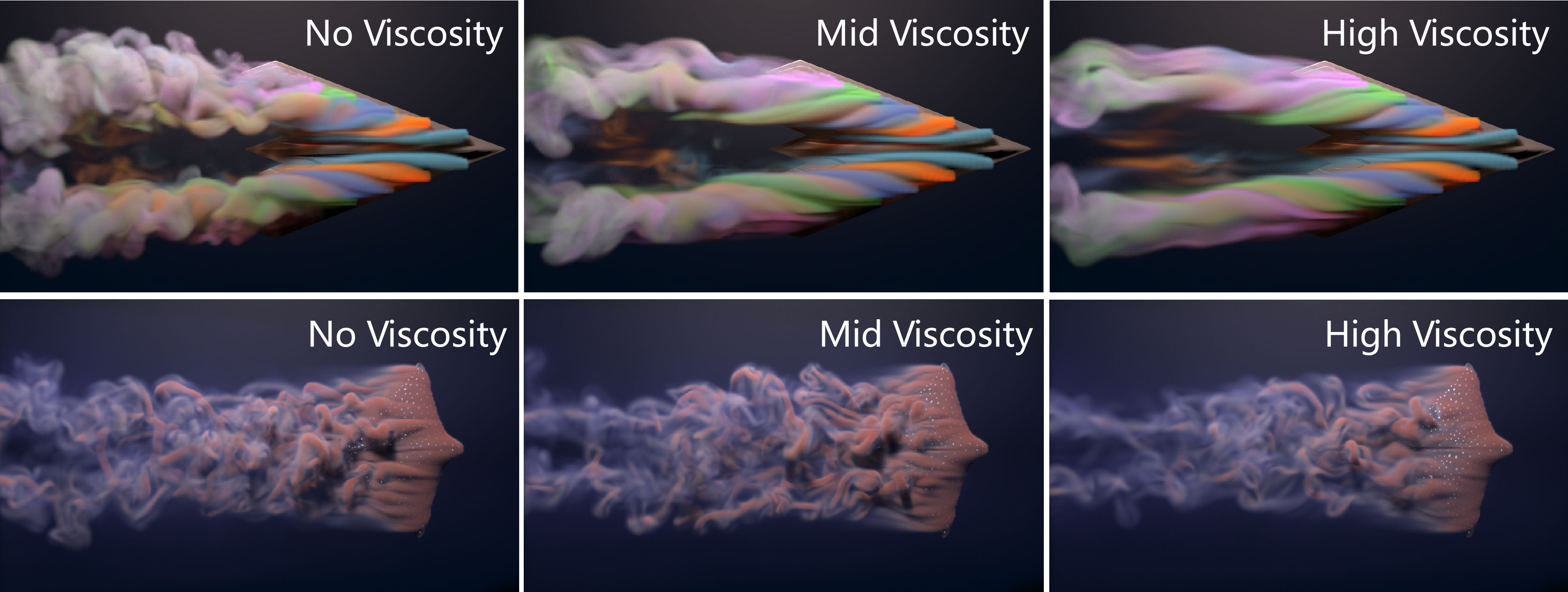}
\captionsetup{aboveskip=6pt}
\caption{
For the top row, we compare delta-wing simulated under different viscosity settings. The left one shows a more turbulent appearance compared to others, while the right one creates stable spiral vortex structures along the two leading edges.
For the bottom row, we show a static stingray countering an incoming flow. A more chaotic and complex structure of vortices can be observed as fluid viscosity reduces.}
\label{fig:ray_and_delta_viscosity_comp}
\vspace{-0.22in}
\end{figure}

\begin{figure*}
\centering
\includegraphics[width=0.995\textwidth]{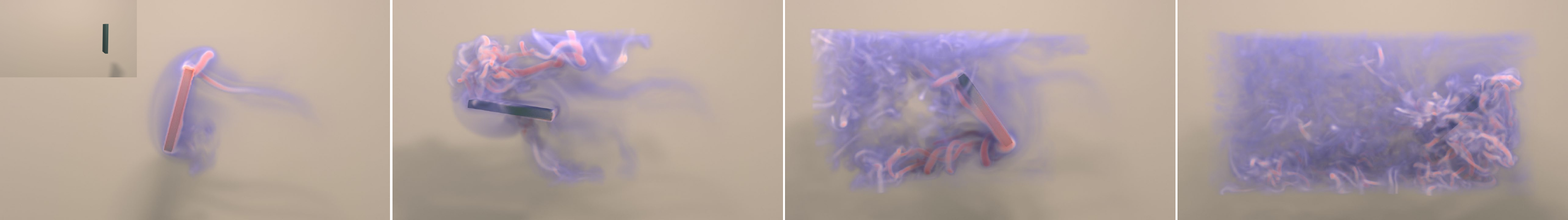}
\captionsetup{aboveskip=6pt}
\caption{
In this scenario, a paddle moves on a preset path inside a box, which results in vortex shedding and creates turbulent flow, particularly around edges and corners. Numerous vortex filaments and tubes are generated during this process.}
\label{fig:paddle}
\end{figure*}

\begin{table*}
\centering\small
\begin{tabularx}{\textwidth}{Y | Y | Y | Y | Y | Y | Y | Y}
\hlineB{3}
\multicolumn{2}{c|}{\textbf{Experimental Settings}} &
\multicolumn{3}{c|}{\textbf{Coupled Solve}} & \multicolumn{3}{c}{\textbf{Separate Solve}}\\
\hlineB{2.5}
Examples & GPU Mem. Provided (GB) & Min Conv. Iters & Max Conv. Iters & Time (sec/solve) & Min Conv. Iters & Max Conv. Iters & Time (sec/solve) \\
\hlineB{2}
Obique & 3 & \textbf{10} & \textbf{10} & \textbf{0.15} & 15 & 114 & 2.40 \\
\hlineB{2}
Obique & 6 & \textbf{10} & \textbf{10} & \textbf{0.14} & 15 & 114 & 2.63 \\
\hlineB{2}
Obique & 9 & \textbf{10} & \textbf{10} & \textbf{0.14} & 15 & 114 & 1.46 \\
\hlineB{2}
Trefoil & 3 & \textbf{10} & \textbf{11} & \textbf{0.16} & 14 & 123 & 1.64 \\
\hlineB{2}
Trefoil & 6 & \textbf{10} & \textbf{11} & \textbf{0.16} & 14 & 123 & 2.22 \\
\hlineB{2}
Trefoil & 9 & \textbf{10} & \textbf{11} & \textbf{0.16} & 14 & 123 & 1.65 \\
\hlineB{2}
Leapfrog & 3 & - & - & - & - & - & - \\
\hlineB{2}
Leapfrog & 6 & 11 & 11 & 0.34 & \textbf{9} & \textbf{10} & \textbf{0.29} \\
\hlineB{2}
Leapfrog & 9 & 11 & 11 & 0.33 & \textbf{9} & \textbf{10} & \textbf{0.28} \\
\hlineB{2}
\end{tabularx}
\vspace{5pt}
\caption{Computational cost comparison between the coupled and separate solutions. The GPU memory (the second column) is set through Taichi \cite{hu2019taichi}. The minimum and the maximum number of iterations of convergence observed during testing are provided. The time column represents the wall-clock time of solving a complete three-dimensional Poisson equation. Note that the 3D Leapfrog experiment could not be executed with 3 GB of GPU memory due to the higher resolution demands.}
\label{tab: couple_table}
\vspace{-0.2in}
\end{table*}

\begin{table}
\centering\small
\begin{tabularx}{0.48\textwidth}{Y  Y  Y }
\hlineB{2}
Name &  Time (sec / frame) & GPU Mem. (GB) \\
\hlineB{1.5}
\textbf{2D Leapfrog}  & 1.69 &  1.54\\
\hlineB{1}
\textbf{3D Leapfrog}  & 2.59 & 5.93 \\
\hlineB{1}
\textbf{3D Trefoil}  & 1.07 & 3.60 \\
\hlineB{1}
\textbf{3D Oblique}  & 0.95 & 3.56 \\

\hlineB{2}
\end{tabularx}
\vspace{5pt}
\caption{Average simulation time per frame and peak GPU memory usage. Each frame typically contains 3-10 steps, with each step involving 2 Poisson solves. Our examples usually consist of 300-600 frames. The CFL number are set to be 1.0 for 2D examples and 0.5 for 3D examples.}
\label{tab:time_memory}
\vspace{-0.2in}
\end{table}

\begin{figure}[t]
\centering
\includegraphics[width=0.48\textwidth]{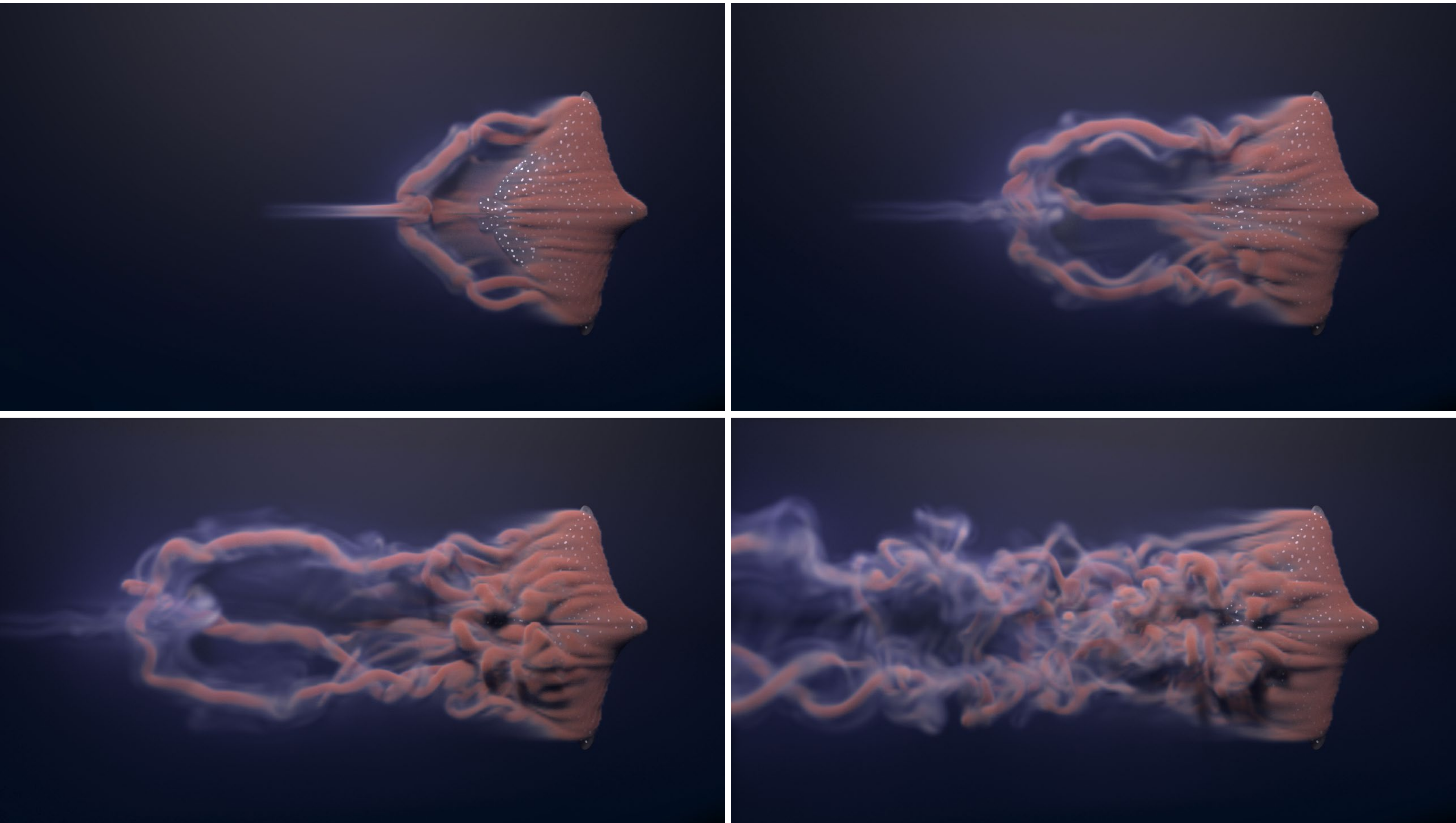}
\captionsetup{aboveskip=6pt}
\caption{A stingray passively glides through water at a constant speed with zero viscosity, as simulated using our Eulerian vortex method. The figure illustrates the turbulent vorticity field generated at different time steps.}
\label{fig:moving_ray}
\end{figure}

\subsection{Performance Analysis and Reinitialization Sensitivity}

\textbf{Performance gains of our coupled approach.}
The performance gains of our coupled approach are provided in Table~\ref{tab: couple_table}. Both the coupled approach and the separate solve runs in parallel. The coupled system is obtained by stacking the separate system along the x-axis, modified with the compatibility condition. As shown in Table~\ref{tab: couple_table}, for oblique vortex collision and trefoil knot, the coupled approach converges faster and requires less time per Poisson solve, achieving speedups of \textbf{10}-\textbf{20}$\times$. For leapfrog, the performance is similar.

\vspace{0.1cm}
\noindent
\textbf{Simulation time and memory.}
In Table~\ref{tab:time_memory}, we present the performance measurements in terms of the average wall-clock time per frame and the maximum GPU memory for our 2D and 3D examples.

\vspace{0.1cm}
\noindent
\textbf{Reinitialization sensitivity.}
The vorticity field at the $150^\text{th}$ frame of the 3D leapfrogging vortices experiment with different reinitialization steps, $n$, is shown in Figure~\ref{fig:n_sense}. The numerical dissipation is significant for very small $n$, such as $n = 1$, leading to poor preservation of vortex structures. Conversely, for much larger $n$, like $n = 50$, while the ability to preserve vortices improves, the distortion of flow map quantities can introduce artifacts and cause the simulation to become unstable, resulting in a higher number of iterations needed for convergence. As shown in the top left image of Figure~\ref{fig:n_sense}, the two vortex rings merged at an early stage due to the large numerical dissipation. In contrast, artifacts appear when the reinitialization step is too large, causing distortion in the flow map quantities.

\begin{figure*}[t]
 \centering
 \includegraphics[width=.99\textwidth]{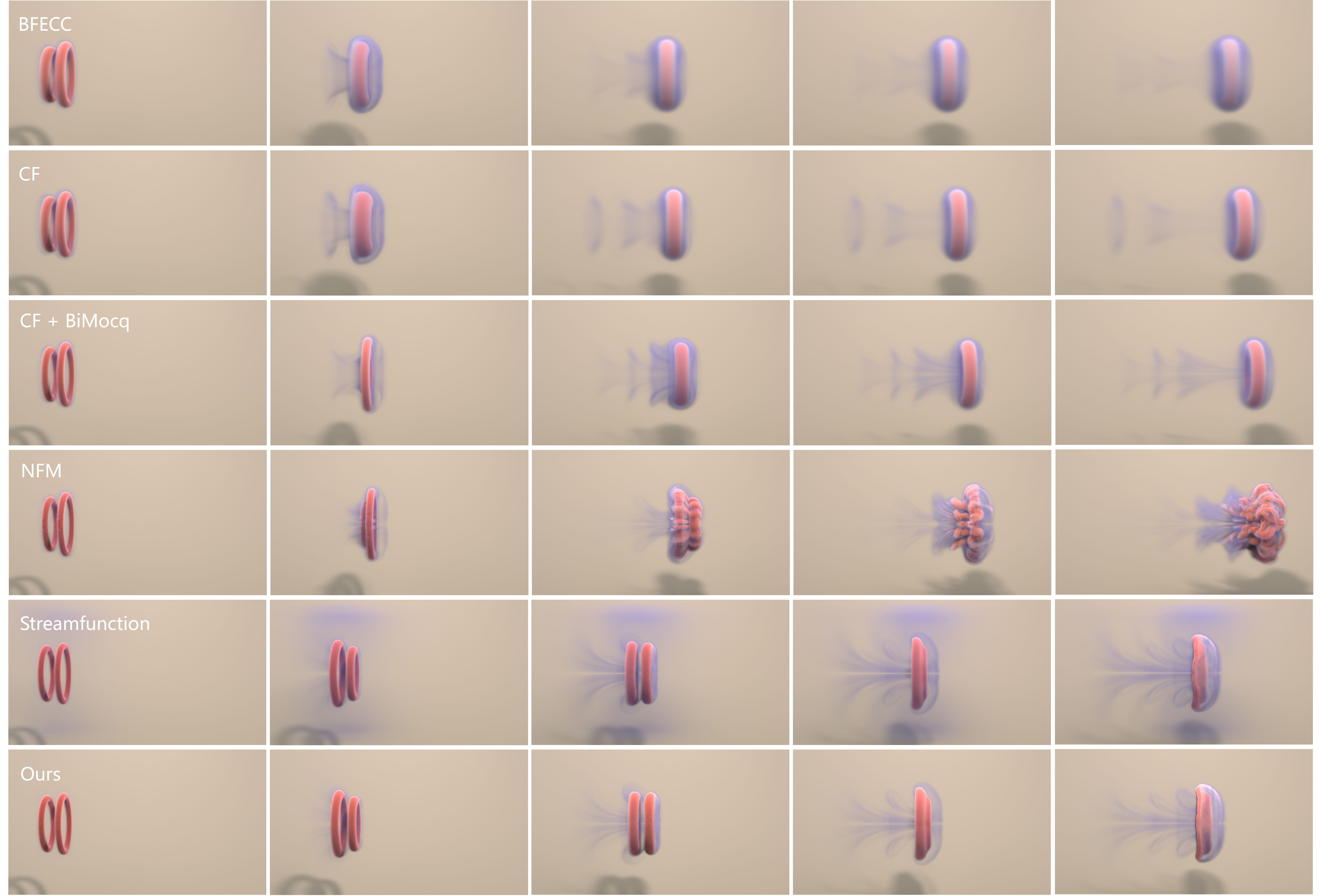}
\captionsetup{aboveskip=8pt}
 \caption{Comparison of 3D leapfrog vortices against benchmarks.}
 \label{fig:3D_leapfrog_compare}
 \vspace{-0.1in}
\end{figure*}

\subsection{2D Examples}
\textbf{2D leapfrog vortices in a box.}
In Figure~\ref{fig:2D_comparison} (top), two vortex pairs released from the left engage in leapfrog-like rotation, moving rightward. Nearing the right wall, they split and return along the top and bottom walls. Colliding with the left wall, they reform and repeat this cycle. Ours maintains vortex separation and structure for over 2000 frames, surpassing others that last less than 400 frames.

\vspace{0.1cm}
\noindent
\textbf{2D Taylor vortex.}
We adopt the setup outlined in \cite{mckenzie2007hola} and \cite{qu2019efficient}, where two vortices are positioned adjacently. Setting the distance at 0.81, our method correctly reproduces the vortex separation phenomenon as shown in Figure~\ref{fig:2D_comparison} (bottom).

\subsection{3D Examples}
\textbf{Head-on vortex collision.}
In Figure~\ref{fig:headon_smoke}, two opposing vortex rings collide and expand. Smaller vortices form along the periphery, mirroring the experimental phenomena in \cite{lim1992instability}.
%
%

\vspace{0.1cm}
\noindent
\textbf{Trefoil knot.}
Using the initialization file from \cite{nabizadeh2022covector}, Figure~\ref{fig:trefoil_knot} shows the knot breaking up and reforming into a larger and a smaller ring, which is in line with physical experiment findings in \cite{kleckner2013creation}.

\vspace{0.1cm}
\noindent
\textbf{3D leapfrog vortices.}
Here, we present the 3D version of the 2D leapfrogging vortices, following the same setup as described in \cite{deng2023fluid}. Initially, two vortex rings are placed separately, moving in the same direction. We compare our results with those of BFECC \cite{kim2006advections}, CF \cite{nabizadeh2022covector}, CF + BiMocq \cite{nabizadeh2022covector, qu2019efficient}, NFM \cite{deng2023fluid}, and the streamfunction-based solver \cite{Ando:2015:streamfunc}, as shown in Figure~\ref{fig:3D_leapfrog_compare}.
The variation using the streamfunction solver differs in the velocity reconstruction step, where the streamfunction solver from \cite{Ando:2015:streamfunc} is applied. To be more specific, the streamfunction is first solved from the vorticity using a Poisson equation, and then the curl of the streamfunction is taken to obtain the velocity. After the fourth leap, both our method and its streamfunction version, as well as NFM \cite{deng2023fluid}, maintain the separation of the two rings, whereas in the other methods, the rings merge before the third leap. The smoke visualization is presented in Figure~\ref{fig:3D_leapfrog}.

\vspace{0.1cm}
\noindent
\textbf{Oblique vortex collision.}
Two perpendicular vortex rings collide, connect, and deform, with the collision point bulging outward. Later, lateral vortices detach, forming a central ring as shown in Figure~\ref{fig:oblique_vort}.

\vspace{0.1cm}
\noindent
\textbf{Moving paddle.}
In Figure~\ref{fig:paddle}, a square paddle in a box moves in a kinematic path, inducing velocity difference around its edges and corners, creating a set of vortex filaments and tubes.

\vspace{0.1cm}
\noindent
\textbf{Delta wing.}
A delta wing featuring a swept angle of \(75^\circ\) is put under an airflow with an angle-of-attack to be \(20^\circ\). The phenomenon of "vortex lift" \cite{anderson2010aircraft} is observed, i.e., spiral vortex sheets and cores appear along the leading edges, as shown in Figure~\ref{fig:delta_wing} and \ref{fig:ray_and_delta_viscosity_comp} (top), which agrees with experimental results in \cite{delery2001robert}.

\vspace{0.1cm}
\noindent
\textbf{Gliding stingray with viscosity.}
Figure~\ref{fig:moving_ray} illustrates a stingray passively gliding, generating vortices. Figure~\ref{fig:ray_and_delta_viscosity_comp} (bottom) displays the stingray with inflow under different viscosity conditions.

\vspace{0.1cm}
\noindent
\textbf{Gliding eagle.}
The eagle is positioned at a \(15^\circ\) angle-of-attack, and generates a rich bundle of vortices from solid-fluid interaction as depicted in the vorticity visualization in Figure~\ref{fig:eagel}.

\section{Conclusion \& Limitation}
This paper presents a novel Eulerian vortex method that combines vortex evolution on flow maps with a carefully devised velocity reconstruction scheme to enable the high-quality simulation of intricate vortical structures and turbulent flows.
Our key insight lies in that, while good results with impulse have been achieved, vorticity promises to work better with flow maps, as it is more numerically stable and physically interpretable.
The efficacy of our method is validated by a number of challenging simulation scenarios.
Our method is subject to a few limitations. First, our method is not able to handle compressible flows. Second, the current system does not support free-surface cases due to nontrivial free-surface boundary treatments required. Extending it to free-surface cases would be a possible and attractive future research direction. Third, while our solver effectively simulates solid-fluid interactions using voxelized boundary models, our system currently does not support boundaries or obstacles with cut cell geometries. Enabling more advanced solid-fluid coupling schemes would be an exciting future direction.
Moreover, our revised Poisson system is symmetric (refer to our solver details in the Appendix Section~\ref{sec:sym} to see that) yet lacks proof for positive definiteness, but the solver converges in all our experiments, with the number of iterations of convergence ranging from 8 to 12 for all examples without solids, and remaining below 50 when complex solid boundaries are present. Unfortunately, we  are still unable to empirically state that all Poisson systems are positive definite, especially given complex solid boundary conditions.
Besides, we tried to model the harmonic component within our solver implicitly, but the problem is not fully solved, as can be seen from the difference between our result and the ground truth in Figure~\ref{fig:harmonic_4}. Resolving the issue of the harmonic component in our method so that it can exhibit behavior identical to that achieved by explicitly modeling the harmonic component will be an interesting direction for future research.
Finally, while our approach mitigates the issue of 
constant impulse increase, which allows for possibly sustaining arbitrarily long flow maps without instability, numerical distortions in the flow map evolution currently necessitate reinitialization, which poses a future challenge to further extend the flow map length.

\section{Acknowledgments}
We sincerely thank the anonymous reviewers for their valuable feedback. We extend our appreciation to Taiyuan Zhang for his assistance with remote access to computer systems, and to Mengdi Wang for his insightful discussions. Georgia Tech authors acknowledge NSF IIS \#2433322, ECCS \#2318814, CAREER \#2433307, IIS \#2106733, OISE \#2433313, and CNS \#1919647 for funding support. The work is also in part supported by Google. We credit the Houdini education license for video animations.

\newpage
\bibliographystyle{ACM-Reference-Format}
\bibliography{refs_ML_sim.bib, refs_INR.bib, refs_flow_map.bib, refs_simulation.bib}

\appendix
\section{Multigrid Solver Details}
\label{sec:app_multigrid}

\subsection{Detailed implementation of the GPU Solver}
\label{sec:app_multi_detail}
The implementation of our MGPCG for our revised Poisson scheme is summarized here and given by Alg.~\ref{alg:mgpcg_main} for easier reproduction:
To setup the equation for an unknown, we need to know how to manipulate the \textbf{RHS} and \textbf{LHS} with the presence of solid boundaries. Suppose we are solving an unknown \(u_x\) , i.e., a x-component of velocity. The initial \textbf{LHS} is
\begin{equation}
    \textbf{NeighborNum}(u_x) \cdot u_x - \textbf{NeighborSum}(u_x)
\end{equation}
where the initial \textbf{NeighborNum}$(u_x)$ is $2 \cdot dim$ ($dim$ is 2 or 3) and \textbf{NeighborSum}$(u_x)$ is the sum of all adjacent velocities to \(u_x\). The initial \textbf{RHS} is the discretized \(\curl \omega\). Then according to the solid boundary conditions, we need to manipulate the \textbf{LHS} (Alg.~\ref{alg:neighbor_num} and Alg.~\ref{alg:neighbor_sum}) and \textbf{RHS} (Alg.~\ref{alg:rhs}) as follows.

In 3D, \(u_x\) will have 6 neighbors (adjacent velocities): 2 each from every axis. We consider these 6 neighbors one by one. The two neighbors along the y or z axis are special because they are not along the same axis as \(u_x\). The key difference lies in that there will be a intervening vorticity variable between \(u_x\) and its y or z axis neighbors. When considering these neighbors, if the intervening vorticity variable is not within fluid (i.e., either on a solid corner or on a solid face or inside the solid), we adjust the LHS in the following way. We subtract one from the coefficient of \(u_x\), remove the corresponding velocity variable from the LHS and add or subtract (depending on the relative position of the unknown and \(u_x\), refer to Section~\ref{sec:sec4.1} to see a concrete example) unknowns within fluids from other axes to the LHS. For the RHS, we remove such intervening vorticity variable, and move the known quantity (those velocities on the walls) to the RHS.
For neighbors along the x axis, if it is on the solid surface, we move it to the RHS.
After configuring the LHS and RHS for each unknown in the system, the system can be solved using a standard conjugate gradient method with a multi-grid preconditioner through a matrix-free approach.

\subsection{Symmetric System}
\label{sec:sym}
It is evident from our algorithm that the system is symmetric. Specifically, whenever we add or subtract an unknown from a different axis, that unknown also reciprocally adds or subtracts the current unknown in its corresponding equation.

\section{Experimental Settings}

Table \ref{tab: examples_table} shows all the experimental settings used in our paper.

\begin{algorithm}
\caption{MGPCG Main}
\label{alg:mgpcg_main}
\begin{flushleft}
         \textbf{Input:} The vorticity field $\bm {\omega_x}$, $\bm {\omega_y}$, $\bm {\omega_z}$
\end{flushleft}
\begin{flushleft}
         \textbf{Output:} The reconstructed velocity field $\bm {u_x}$, $\bm {u_y}$, $\bm {u_z}$
\end{flushleft}
\begin{algorithmic}[1]
\State $\bm r_x \gets \curl \bm {\omega_x}$; $\bm r_y \gets \curl \bm {\omega_y}$; $\bm r_z \gets \curl \bm {\omega_z}$;
\State Stack \(\bm {r_x}, \bm {r_y}, \bm {r_z}\) as \(\bm r\) along the x-axis.
\State Initialize a system of equations $\bm {\mathcal{S}}$ with the same shape of \(\bm r\)
\State \textbf{RHS}$(\bm {\mathcal{S}}) \gets$ \textbf{SetUpRHS}\((\bm r)\) according to Alg.~\ref{alg:rhs}.

\ForAll{grid position $\bm I = (i, j, k)$ \textbf{in parallel}}
\If{$u$ is within fluid}
\State Setup \(\textbf{LHS}(u)\) according to Alg.~\ref{alg:neighbor_num} and Alg.~\ref{alg:neighbor_sum}.
\EndIf
\EndFor
\State $\bm {u_x}$, $\bm {u_y}$, $\bm {u_z} \gets$ \textbf{MGPCG\_Solve}\((\bm{\mathcal{S}})\)
\end{algorithmic}
\end{algorithm}

\begin{algorithm}
\caption{SetUpRHS}
\label{alg:rhs}
\begin{flushleft}
         \textbf{Input:} The stacked $\bm{r}$, the solid boundary $\bm{b}$, the solid velocity $\bm{bv}$
\end{flushleft}
\begin{algorithmic}[1]
\ForAll{grid position $\bm I = (i, j, k)$ \textbf{in parallel}}
\If{$u$ is within fluid}
\State $ca \gets I(u)[0]\mathbin{//} res_x$
\Comment{The current solving axis.}

\For{$a$ in $axes$}
\If{$a$ = $ca$}
\State Add neighboring velocities to $\bm {r}(u)$ if on solids
\Else
\If{Neighboring vorticities not within fluids}
\State remove the neighboring vorticies from $\bm {r}(u)$
\EndIf

\EndIf
\EndFor
\EndIf
\EndFor
\end{algorithmic}
\end{algorithm}

\newpage

\begin{algorithm}
\caption{NeighborSum}
\label{alg:neighbor_sum}
\begin{flushleft}
         \textbf{Input:} the velocity unknown $u$, its grid position $I(u)$, the solid boundary $\bm{b}$
\end{flushleft}
\begin{flushleft}
         \textbf{Output:} the sum of the neighbors $\bm s$
\end{flushleft}
\begin{flushleft}
         \textbf{Initialize:} $\bm s$ to 0
\end{flushleft}
\begin{algorithmic}[1]

\If{$u$ is within fluid}
\State $ca \gets I(u)[0]\mathbin{//} res_x$
\Comment{The current solving axis.}

\For{each axis $a$}
\If{$a$ != $ca$}
\If{the neighboring vorticity is not within fluids}
\State Add/subtract velocities within fluids from $a$ to $\bm s$
\Else
\State Add neighboring velocities within fluids to $\bm s$ 
\EndIf
\Else
\State Add neighboring velocities within fluids to $\bm s$ 

\EndIf
\EndFor

\EndIf
\end{algorithmic}
\end{algorithm}

\begin{algorithm}
\caption{NeighborNum}
\label{alg:neighbor_num}
\begin{flushleft}
         \textbf{Input:} the velocity unknown $u$, its grid position $I(u)$, the solid boundary $\bm{b}$, the solid velocity $\bm{bv}$
\end{flushleft}
\begin{flushleft}
         \textbf{Output:} the coefficient $\bm n$ of $u$
\end{flushleft}
\begin{flushleft}
         \textbf{Initialize:} $\bm n$ to $2 \cdot dimension$
\end{flushleft}
\begin{algorithmic}[1]
\If{$u$ is within fluid}
\State $ca \gets I(u)[0]\mathbin{//} res_x$
\Comment{The current solving axis.}

\For{each axis $a$ != $ca$}
\For{each neighboring vorticity not within fluids}
\State $\bm n -= 1$
\EndFor
\EndFor

\EndIf
\end{algorithmic}
\end{algorithm}

\begin{table*}[t]
\centering\small
\begin{tabularx}{\textwidth}{Y | Y | Y | Y | Y}
\hlineB{3}
Name  & Resolution & CFL & Reinit Steps & Inflow velocity\\
\hlineB{2.5}
2D leapfrog & 256 $\times$ 256 & 1.0 & 20 & -\\

\hlineB{2.5}
2D Taylor vortex & 256 $\times$ 256 & 1.0 & 20  & -\\

\hlineB{2.5}
2D von Kármán vortex street & 256 $\times$ 512 & 1.0 & 20  & 0.16\\

\hlineB{2.5}
2D Lid-driven cavity flow & 256 $\times$ 256 & 1.0 & 20 & 1\\

\hlineB{2.5}
2D vortex pair passing disks & 512 $\times$ 512 & 1.0 & 20 & -\\

\hlineB{2}
3D leapfrog  & 256 $\times$ 128 $\times$ 128 & 0.5 & 20 & -\\

\hlineB{2}
3D oblique  &128 $\times$ 128 $\times$ 128 & 0.5 & 10 & -\\

\hlineB{2}
3D headon  & 128 $\times$ 256 $\times$ 256 & 0.5 & 16  & -\\

\hlineB{2}
3D trefoil  & 128 $\times$ 128 $\times$ 128 & 0.5 & 10 & -\\

\hlineB{2}
3D moving paddle  & 256 $\times$ 128 $\times$ 128 & 0.5 & 8  & -\\

\hlineB{2}
3D delta-wing  & 256 $\times$ 128 $\times$ 128 & 0.5 & 12 & 0.1 \\

\hlineB{2}
3D eagle & 256 $\times$ 256 $\times$ 128 & 0.5 & 12 & 0.1\\

\hlineB{2}
3D stingray  & 256 $\times$ 128 $\times$ 128 & 0.5 & 12 & 0.1\\

\hlineB{3}
\end{tabularx}
\vspace{5pt}
\caption{All experimental settings included in the paper.}
\label{tab: examples_table}
\end{table*}

\end{document}